\documentclass[times]{aastex631}
\usepackage{amssymb,amsmath,mathrsfs,mathtools,amsthm}
\newtheorem{lem}{Lemma}
\newtheorem{prop}{Proposition}
\usepackage{rotating}

\newcommand{\mean}[1]{\text{E} \left [ #1 \right ]}

\newcommand{\var}[1]{\text{Var} \left [ #1 \right ]}

\newcommand{\trace}{\text{tr}}
\newcommand{\vect}[1]{\boldsymbol{#1}}
\newcommand{\matr}[1]{\mathbf{#1}}
\newcommand{\diag}{\text{diag}}
\newcommand{\df}{\text{df}}
\newcommand{\tp}{^\top}

\received{July 25, 2021}
\revised{July 25, 2021}
\accepted{\today}

\submitjournal{ApJ}

\shorttitle{
Periodic Variable Stars Modulated by Time-Varying Parameters}
\shortauthors{Motta et al.}

\begin{document}

\title{Periodic Variable Stars Modulated by Time-Varying Parameters \footnote{Released on July, 25th, 2021}}

\correspondingauthor{Darlin Soto}
\email{dmsoto1@uc.cl}

\author{Giovanni Motta}
\affiliation{Department of Statistics, Texas A\&M University, 3143 TAMU, 155 Ireland Street, College Station, TX 77843-3143, USA}
\author{Darlin Soto}
\affiliation{Department of Statistics, Faculty of Mathematics,
Pontificia Universidad Cat\'{o}lica de Chile, 7820436 Macul, Santiago, Chile}
\author[0000-0001-6003-8877]{M\'{a}rcio Catelan}
\affiliation{Instituto de Astrof\'{i}sica, Facultad de F\'isica, Pontificia Universidad Cat\'{o}lica de Chile, 7820436 Macul, Santiago, Chile}
\affiliation{Centro de Astro-Ingeniería, Pontificia Universidad Cat\'{o}lica de Chile, 7820436 Macul, Santiago, Chile}
\affiliation{Millennium Institute of Astrophysics, Nuncio Monse\~{n}or Sotero Sanz 100, 7500000, Santiago, Chile}

\begin{abstract}

Many astrophysical phenomena are time-varying, in the sense that their brightness change over time. In the case of periodic stars, previous approaches assumed that changes in period, amplitude, and phase are well described by either parametric or piecewise-constant functions. With this paper, we introduce a new mathematical model for the description of the so-called modulated light curves, as found in periodic variable stars that exhibit smoothly time-varying parameters such as amplitude, frequency, and/or phase. Our model accounts for a smoothly time-varying trend, and a harmonic sum with smoothly time-varying weights. In this sense, our approach is flexible because it avoids restrictive assumptions (parametric or piecewise-constant) about the functional form of trend and amplitudes. We apply our methodology to the light curve of a pulsating RR Lyrae star characterised by the Blazhko effect. To estimate the time-varying parameters of our model, we develop a semi-parametric method for unequally spaced time series. The estimation of our time-varying curves translates into the estimation of time-invariant parameters that can be performed by ordinary least-squares, with the following two advantages: modeling and forecasting can be implemented in a parametric fashion, and we are able to cope with missing observations. To detect serial correlation in the residuals of our fitted model, we derive the mathematical definition of the spectral density for unequally spaced time series. The proposed method is designed to estimate smoothly time-varying trend and amplitudes, as well as the spectral density function of the errors. We provide simulation results and applications to real data.
\end{abstract}


\section{Introduction}\label{sec:intro}
RR Lyrae stars are important astrophysical tools for the measurement of distances and studies of the astrophysical properties of old stellar populations. They are moderately bright, evolved low-mass stars, currently in the core helium-burning phase, also known as the horizontal branch. Their periods are typically in the range between about 0.2 and 1.0~d, which together with their characteristic light-curve shapes, allow them to be relatively easily identified in time-series photometric surveys. An overview of their properties can be found in the monographs by \citet{hs95} and \citet{cs15}.

In spite of their astrophysical importance, RR Lyrae stars are still not fully understood. Indeed, one of the longest-standing problems in stellar astrophysics is also one that specifically affects RR Lyrae stars: the so-called {\em Blazhko effect} \citep{sb07}. It consists in a long-term modulation of an RR Lyrae's light curve, over timescales ranging from a few to hundreds of days \citep[for recent reviews, see][]{cs15,gk16}. The Blazhko effect is particularly common amongst fundamental-mode (ab-type) pulsators \citep[e.g.,][]{epea19}, but is also present, to a lesser extent, in first-overtone (c-type) RR Lyrae stars \citep[e.g.,][]{hnea18}.

Over the decades since it was first described, the Blazhko effect has persistently defied theoretical explanations as to its cause \citep[e.g.,][]{gk16}. Gradual strengthening and weakening of turbulent convection in the stellar envelope \citep{rs06}, a 9:2 resonance between the fundamental and ninth-overtone radial modes \citep{bk11}, and interaction between fundamental and first-overtone modes in the ``either-or'' region of the instability strip \citep{dg13} are the most recent candidates, but no consensus has yet been reached as to the root cause of the Blazhko effect, due in large part to the difficulties involved in the non-linear hydrodynamical modeling of the phenomenon.

In this article, we introduce a model for time series observations of variable stars having smoothly time-varying trend and amplitudes. More precisely, we develop a semi-parametric method for unequally spaced  time series measuring the  brightness of a modulated variable star. Our approach is flexible because it avoids assumptions about the functional form of trend and amplitudes. The estimation of our time-varying curves translates into the estimation of time-invariant parameters that can be performed by ordinary least-squares, with the following two advantages: modeling and forecasting can be implemented in a parametric fashion, and we are able to cope with missing observations. We also study the spectral density of the residuals obtained from the fit of our novel model. 

In order to detect serial correlation in the residuals, in this paper we derive the definition of the spectral density for unequally spaced time series.  There are many reasons why astronomical time series are not sampled equidistantly, and the gaps can be either regular or random. From the Earth, stars can't be observed during the day, which introduces regular gaps in the time series. Also, for about half a year, most objects become unobservable, as they are up on the sky at the same time as the Sun, which introduces yearly gaps. There could be clouds or high wind, forcing the closure of telescopes, producing random gaps. There could be high-priority alerts overriding the observations, or the telescope could be available only on certain nights.

In some cases, observations are unevenly spaced due to missing values. Astronomical data sets often contain missing values, and this limitation is sometimes due to incomplete observations or varying survey depths. Even if telescopes are recording and storing information systematically, that is, at a regular cadence, there are a few things that can 
alter the regular sampling. For example, an astronomer might decide to increase the exposure time if there are clouds obscuring the target, to try to increase the signal-to-noise ratio. Conversely, if the observing conditions are excellent, the astronomer might decide to decrease the exposure times (and hence the cadence) to avoid saturating the detector, for example. Missing values are usually handled via imputation, that is, the gap generated by the missing value is ``filled in" by an estimated value. If observations are missing because of the survey, imputation can be performed using statistical models. For example, \cite{FBC2018} apply ARIMA models to fill in missing data in astronomical time series. 

However, in astrostatistics, missing value problems are sometimes inherently brought about by the manner in which  physical processes are recorded. In particular, telescopes are not located in the center of the Solar System. Since the speed of light is finite, this results in a time delay between the arrival times of signals at our position and at the center of the Solar System. This is typically corrected for by referring the times of observations to either Heliocentric Julian Dates (HJD) or Barycentric Julian Dates (BJD), which refer to the center of the Sun or the entire Solar System, respectively. Thus, even if telescopes record data strictly evenly according to the local time at the observatory (e.g., one observation performed every night at local midnight), this correction will slowly change between observations, modifying what was initially a regular grid to an irregular one. Also, this correction is different for every source on the sky, even though sources close to each other may have very similar corrections. Therefore, for some astronomical data sets where missing values may arise from the manner in which observations of a physical process are collected, or even the nature of the physical process itself (e.g., sudden, extreme dimming events that may occasionally render an object impossible to detect for a certain amount of time), the imputation method may not be applicable \citep[see][]{Chattopadhyay2017}. Our novel approach, which involves the classical periodogram, has the advantage of not relying on any imputation method.

We shall divide the present study into six main sections. In Section~\ref{Sec:model} we introduce our novel model and clarify analogies and differences as compared with previous approaches. In Section~\ref{Sec:modulation}, the present status of important ingredients of amplitude and frequency modulations is critically discussed. In Section~\ref{Sec:estimation} we present the method we adopt to estimate the time-varying parameters. In Section~\ref{Sec:corr} we present a new method to estimate the spectral density of unequally spaced times series, which is needed for the analysis of the residuals. Section~\ref{Sec:simres} provides simulation results, whereas Section~\ref{Sec:appl} illustrates the advantages of using our novel method by means of an application to an RR Lyrae variable star. Finally, our main conclusions are summarized in Section~\ref{Sec:summary}.

Through the paper we use bold uppercase letters to denote matrices, and bold slanted to denote vectors. We denote by $\mathbf I_m$ the identity matrix of size $m$, by $\vect{0}_{n}$ a column-vector of zeros of length $n$, by ${\rm tr}\{ \mathbf A\}$ the trace of $\mathbf A$, 
by $\mathbf A\tp$ the transpose of $\mathbf A$, by $\|\mathbf A\|$ the Frobenius norm $\|\mathbf A\|=[{\rm tr}\{ \mathbf A\tp \mathbf A\}]^{\scalebox{0.6}{$1/2$}}$, and by $\mathbf A^{-1}$ the inverse of the square matrix $\mathbf A$, that is, the square matrix $\mathbf A^{-1}$ such that $\mathbf A^{-1}\mathbf A=\mathbf A\mathbf A^{-1}=\mathbf I$.

\section{A NOVEL TIME-VARYING MODULATION-MODEL FOR VARIABLE STARS}\label{Sec:model}
Light curves of variable stars are typically fitted using harmonic models with linear (or constant)  trend and time-invariant amplitudes \citep[see equations (1) and (5) in ][]{Richards_2011}. This type of model would be inappropriate when the underlying trend and amplitudes change over time in a more complex way. \cite{Eilers2008} proposed a model with one harmonic component ($K=1$) where trend and amplitudes vary smoothly over time. In this paper, we extend the model by \cite{Eilers2008} to  the case of $K\geq 1$ harmonic components, where the amplitudes associated with each harmonic component vary smoothly over time. We estimate our model by means of P-splines \citep{Eilers1996}, which are a combination of $B$-splines and penalties. The estimation of the time-varying curves translates into the estimation of time-invariant parameters that can be performed by the least squares method, with the following three advantages: it is computationally fast, forecasting can be implemented in a parametric fashion, and we can cope with missing observations.

Compared to local smoothers (such as kernel smoothers), the main advantage of regression spline in the context of time series is that the unknown parameters are time-invariant and thus they can be estimated globally rather than locally. As a consequence, forecasting only requires good estimates of the global unknown parameters. We can think of a regression splines with $B$-splines as a semi-parametric model in the sense that it contains parametric as well as non-parametric components. The parametric component is given by a finite number of parameters, whereas the non-parametric component by the basis functions. 
Another advantage of parametric and semi-parametric models over non-parametric models is the computing speed, as many non-parametric models are computationally intensive. Finally, the use of $B$-splines in regression allows us to rewrite the estimation problem as a least squares fit, avoiding the use of numerical methods -- such as Newton Raphson -- which can be time consuming.

Let $\{Y_{\!i}\equiv Y_{\!{t_i}},\, i=1,\dots,N\}$ be a set of observations occurring at certain discrete times $t_1,\dots,t_N $. In the case of equally spaced observations, $t_i=t_0+i\Delta$ where $i$ is an integer, and $\Delta>0$ is the constant data spacing. Then  $|t_i-t_k|=\Delta|i-k|$, and typically $\Delta=1$. Astronomical light curves are often observed unequally in time, that is, the data spacing of observation times is not constant. We decompose the observed light curve into the sum of a deterministic periodic trend and a random noise as

We decompose the observed light curve into the sum of a deterministic signal $\mu$ and a random noise $z$. The deterministic part $\mu(t)$ consists of a trend $m(t)$ and a {\it modulated} periodic signal. The modulated periodic signal is a linear combination of $K$ cosines and sines, with weights given by the {\it modulating} functions $g(t)$:
\begin{equation}\label{mod_model}
\begin{split}
Y_{\!_{i}} &=\mu(t_i)+z_{\!_{i}}, \qquad i=1,\dots,N, \qquad \{z_{\!_{i}}\} \overset{}{\sim} WN(0,\sigma^2_{\!{z}}),\\
\mu(t_i) &=m(t_i)+\sum^K_{k=1}\{g_{\!_{1,k}}(t_i)\cos(w_{\!_{k}}t_i)+g_{\!_{2,k}}(t_i)\sin(w_{\!_{k}}t_i)\},
\end{split}
\end{equation}
or in matrix notation $\vect{Y}=\vect{\mu}+\vect{z}$,
where $\vect{Y}=(Y_{\!_{1}},\dots,Y_{\!_{N}})\tp$ is the vector of observations at time $\vect{t}=(t_{\!_1},\dots, t_{\!_N})\tp$, $\vect{\mu}=[\mu(t_1),\dots,\mu(t_N)]\tp$ is the expectation of $\vect{Y}$, $m(t_i)$ is the smooth time-varying trend at time $t_i$, the $g(t_i)$'s are smooth time-varying amplitudes of the cosine and sine waves at time $t_i$, respectively, $w_{\!_{k}}=2\pi f_{\!_{k}}$ is the angular frequency, and $f_{\!_{k}}$ is the ordinary frequency. Since the errors are zero-mean, the expectation of the observed brightness at time $t_i$ is equal to the deterministic part of the signal at time $t_i$, that is, $\mean{Y_i}=\mu(t_i)$.

We refer to $m(\cdot)$ as the ``trend", that is, the (typically) aperiodic change in the mean of the light
curve. On the other hand, we call ``amplitudes" the functions $g(\cdot)$'s that weigh the periodic variation (of this average brightness) of cosine and sine waves.
In Appendix~\ref{Ape_modulation} we summarize the standard mathematical definitions of amplitude modulation and frequency modulation. Both trend and amplitudes are typically restricted to be sinusoidal, whereas in this paper our trend $m(t)$ and our amplitude functions $g(t)$'s are general smooth functions and not necessarily sinusoidal. In Section~\ref{Sec:modulation} we clarify the mathematical connection between the standard modulation models and our novel modulation model in equation  \eqref{mod_model}.

The error vector $\vect{ z}=(z_{\!_{1}},\dots,z_{\!_{N}})\tp$ is a \textit{white noise} (WN) process with mean zero and variance $\sigma^2_{\!{z}}$. That is, each error $z_{\!_{i}}$, $i=1,\dots,N$, follows a zero-mean WN process with variance $\sigma^2_{\!{z}}$:
\[
\mean{z_{\!_{i}}}=0\quad\mbox{and} \quad \mean{z_{\!_{i}}\,z_{\!_{j}}}=\delta_{\!_{\{i=j\}}}\sigma^2_{\!{z}},\quad\mbox{for all $i,j=1,\dots,N$},
\]
where $\delta_{\!_{\{i=j\}}}=1$ if $i=j$ and zero otherwise. 

Our model in equation \eqref{mod_model} is defined in discrete time, and it focuses on time-domain. \cite{Kelly2014}  adopt the continuous-time auto-regressive moving average (CARMA) models to estimate the variability features of a light curve in the frequency domain. More specifically, \cite{Kelly2014} use the power spectral density (PSD) of CARMA models to account for irregular sampling and measurement errors. A stationary CARMA($p, q$) process has the PSD
\[
P(f ) = \sigma^2 {\big|\displaystyle\sum_{j=0}^q\beta_j (i 2 \pi f)^j\big|^2}/
{\big|\displaystyle\sum_{k=0}^p\alpha_k (i 2 \pi f)^k\big|^2}.
\]
To illustrate the importance of fitting models with time-varying parameters, \cite{Kelly2014} simulated a light curve that switches from one CARMA process to another. More precisely, they constructed a non-stationary light curve by generating two CARMA processes of the same order ($p=5, q=3$), but with different parameters \citep[see][Section 4.3]{Kelly2014}:
\[\boldsymbol\theta(t)=
\begin{cases}
\boldsymbol\theta_1& t_1\leq t < t_0\\
\boldsymbol\theta_2& t_0\leq t\leq t_N,
\end{cases}
\]
where $\boldsymbol\theta(t)=[\alpha_1(t),\dots,\alpha_p(t),\beta_1(t),\dots,\beta_q(t),\sigma^2(t)]^\top$. The vector  $\boldsymbol\theta(t)$ is a step-wise function that is constant before and after $t_0$. The approach based on piece-wise constant parameters is receiving growing interest in various areas of astrophysics. \cite{Wong2016} adopt a Poisson model for the photon counts. They define $\lambda(t_j,w_i)$ as the expected
count per unit time and per unit wavelength averaged over the bin centered at $(t_j,w_i)$, and detect change-points $\pi$ such that $\{\lambda(t_j,w_i)|t_j\leq\pi\}\neq \{\lambda(t_j,w_i)|t_j>\pi\}$. \cite{Wong2016} estimate the number of change points and their values. \cite{Xu2021} develop a method for modeling a time series of
images, and assume that the arrival times of the photons follow a Poisson process. They assume that all image stacks between
any two adjacent change points (in time domain) share the same unknown piece-wise constant function. \cite{Xu2021} estimate the number and the locations of all of the change points (in time domain), as well as all of the unknown piece-wise constant functions between any pairs of the change points.

Instead of considering parameters that are piece-wise constant functions of time, in this paper we allow the parameters to be {\it smooth} functions of \textit{rescaled time}, permitting the process to be {\it locally stationary}. The framework of {local stationarity} introduced by \cite{D97}, where the parameter curves are defined in rescaled time $u=t/T$, provides a meaningful asymptotic theory. Locally stationary $Y(t)$ means that if the functions $m(u)$ and $g(u)$ in equation \eqref{mod_model} are ``smooth" and $T$ is large, $m(\tfrac t T)\approx m(\tfrac r T)$ and $g(\tfrac t T)\approx g(\tfrac r T)$ for values of $r$ close to $t$, that is, {\it locally} around $t$. More precisely, we assume that the functions $m(x)$ and $g(x)$ are Lipschitz continuous, that is, there exist  constants $C_m$ and $C_g$  such that 
\begin{equation}\label{lip}
|m(z)-m(u)|\leq C_m|z-u| \qquad\mbox{and\qquad}|g(z)-g(u)|\leq C_g|z-u|,\qquad\mbox{for all $u,z\in[0,1]$}.
\end{equation}
To define $m(u)$ and $g(u)$ as functions of rescaled time let us consider, for each fixed $u\in[0,1]$ and incresing $T$, the sequence $t=t_T=\lfloor u\,T\rfloor$, where $\lfloor x\rfloor$ is the largest integer not exceeding $x$. Then we obtain the following uniform bound: $\sup_t |\tfrac t T - u|<\tfrac 1 T$. 
A time series is stationary if the moments of the underlying stochastic process, such as expectation and variance,  are time-invariant. The idea behind local stationarity is to allow for time-varying parameters, in a way that {\it locally} the process behaves as stationary. Lipschitz continuity is a smoothness assumption that implies uniform continuity. The model in equations \eqref{mod_model}-\eqref{lip} is a locally stationary process written in rescaled time in a way that, as $T$ grows we observe more and more ``observations" of the same type around $u$. That is, if $m(\cdot)$ and $g(\cdot)$ are smooth we have 
\begin{equation*}
|m(\tfrac{t}{T})-m(u)|\leq C_m|\tfrac{t}{T}-u|<\tfrac{C_m}{T}\to 0\quad\mbox{and\qquad}|g(\tfrac{t}{T})-g(u)|\leq C_g|\tfrac{t}{T} -u|<\tfrac{C_g}{T}\to 0\quad\mbox{as $T\to\infty$},
\end{equation*}
for all $t:= \lfloor u\,T\rfloor$. The locally stationary framework is important for handling, in a meaningful way, the asymptotic theory arising in statistics for processes with time-varying parameters.  Suppose that we observe  $X_t = \mu(t)+ z_t$, with $z_t\sim WN(0, \sigma_z^2)$ for $t = 1,\dots,T$. Inference in this case means studying the properties of an estimator for the unknown function  $\mu(t)$ on the grid $\{1,\dots,T\}$. Given that $\mu$ changes over time, it is obvious that an asymptotic approach where $T\to\infty$ is not suitable for describing a statistical method, since future ``observations" $\{\mu(t),\,t>T\}$ do not necessarily contain any information on $\mu(t)$ on $\{1, \dots,T\}$. To overcome these problems, \cite{D96} suggested to consider a triangular array of data. In analogy with non-parametric regression, it seems natural to set down the asymptotic theory in a way that we ``observe" $\mu(t)$ on a finer grid (but on the same interval), i.e. that we observe the process $Y_T(t)=\mu(\tfrac{t}{T})+z_t$, where $Y_T$ is now a triangular array and $\mu$ is now rescaled to the interval $[0, 1]$. Working in rescaled time is often adopted also within the estimation framework of regression splines \citep[see][among others]{Zhou1998}.

Time series analysis of non-stationary sequences can be deterministic or stochastic. A popular example of stochastic non-stationarity is the  well known class of integrated processes, where the observed times series can be made stationary by differencing. \cite{FBC2018} apply autoregressive integrated moving average (ARIMA) models to light curves of several variable stars, discussing their effectiveness for different temporal characteristics. The process $\{X_t\}$ is an ARIMA($p,d,q$) process if $Y_t=(1-B)^dX_t$, obtained by applying the operator $1-B$ repeatedly $d$ times, is a stationary ARMA($p,q$) process. The most popular example of ARIMA($p,d,q$) process is the ``random walk" $X_t=X_{t-1}+ Z_t$, which is an ARIMA with $p=q=0$  and $d=1$.

In the next section, we review the models proposed by \cite{Benko2011} and \cite{Benko2018} for Blazhko light curves. Interestingly, our model in equation \eqref{mod_model} generalizes the models by \cite{Benko2011} and \cite{Benko2018} in the sense that the modulating functions $g(\cdot)$ are not confined to the class of parametric (sinusoidal or non-sinusoidal) functions.

\section{MODELING BLAZHKO LIGHT CURVES} \label{Sec:modulation}
The Blazhko effect is a periodic amplitude and phase variation in the light curves of RR Lyrae variable stars. In astronomy, the Blazhko effect is usually interpreted as a modulation phenomenon. Modulation is the process of transmitting a low-frequency signal into a high-frequency wave, called the carrier wave, by changing its amplitude, frequency, or phase angle through the modulating signal. In Appendix~\ref{Ape_modulation} we review the main mathematical definitions underlying the modulation phenomenon in astrophysics.

In this section, we review the models proposed by \cite{Benko2011} and  \cite{Benko2018}, respectively, and we compare them with our novel model in equation \eqref{mod_model}. To describe Blazhko light curves, \cite{Benko2011}  proposed to fit the following model:
\begin{equation}\label{Benko_model0}
\mu^*(t)=a^A_{\!_{0}} a_{\!_{0}} + a_{\!_{0}} g^A(t) + \sum^{K}_{k=1} \left[ a^A_{\!_{0}} a_{\!_{k}} + a_{\!_{k}} g^A(t) \right] \sin [2 \pi k f_{\!_{0}} t + \varphi_{\!_{k}} + k g^F(t)],
\end{equation}
where $a_k$ and $f_0$ denote amplitude and frequency, respectively, and
\begin{equation}\label{gMt1}
g^M(t)=\sum^{\ell^M}_{j=1} a^M_{\!_{j}} \sin (2 \pi j f_{\!_{m}} t + \varphi^M_{\!_{j}}), \quad M=A \text{ or } F.
\end{equation}
More recently, \cite{Benko2018} introduced a similar model:
\begin{equation}\label{Benko_model}
\mu^*(t)=m_{\!_{0}} + \sum^{\ell}_{r=1} b_{\!_{r}} \sin(2 \pi r f_{\!_{m}}t+ \varphi^b_{\!_{r}})+\sum^{K}_{k=1} \left [ a_{\!_{k}} + g^A_{\!_{k}}(t) \right ] \sin [2 \pi k f_{\!_{0}} t + \varphi_{\!_{k}} + g^F_{\!_{k}}(t)],
\end{equation}
where $b_r$ and $\phi_r^b$ denote amplitude and frequency of the modulating signal, respectively, and
\begin{equation}\label{gMt2}
g^M_{\!_{k}}(t)=\sum^{\ell^M_{\!_{k}}}_{j=1} a^M_{\!_{kj}} \sin (2 \pi j f_{\!_{m}} t + \varphi^M_{\!_{kj}}), \quad M=A \text{ or } F.
\end{equation}
The functions $g^M(t)$ and $g^M_{\!_{k}}(t)$ in equations \eqref{gMt1} and \eqref{gMt2} are the modulating functions with subscripts $M=A$ and $M=F$ denoting amplitude and frequency modulation, respectively. The main pulsation frequency is denoted by $f_{\!_{0}}$, whereas  $f_{\!_{m}}$ is the modulating frequency. In this paper we improve the models in equations \eqref{Benko_model0}-\eqref{gMt1} and \eqref{Benko_model}-\eqref{gMt2} from two different viewpoints. From the modeling viewpoint, we relax the assumption of parametric amplitude and frequency modulations. Assuming parametric amplitude and frequency modulations results in a Fourier sum with time-invariant amplitudes and time-invariant frequencies, whereas our time-varying amplitudes and frequencies do not obey any particular form. From the estimation viewpoint, we do not rely on the non-linear least squares algorithms, such as the Levenberg-Marquardt algorithm, that are typically used to fit parametric non-linear models. These methods require initial values close to the solution, which in some applications are difficult to find.

Both models proposed by \cite{Benko2011} and \cite{Benko2018} and given by equations \eqref{Benko_model0} and \eqref{Benko_model}, respectively, are a special case of our model defined by equation \eqref{mod_model}. To see this, let us define
\begin{align}\label{Benko_comp2011}
\begin{split}
v(t)&=a^A_{\!_{0}} a_{\!_{0}} + a_{\!_{0}} g^A(t), \\
w_{\!_{1,k}}(t)&= \left[ a^A_{\!_{0}} a_{\!_{k}} + a_{\!_{k}} g^A(t) \right] \sin [\varphi_{\!_{k}} + k g^F(t)], \quad k=1,\dots,K,\\
w_{\!_{2,k}}(t)&=\left[ a^A_{\!_{0}} a_{\!_{k}} + a_{\!_{k}} g^A(t) \right] \cos [\varphi_{\!_{k}} + k g^F(t)], \quad k=1,\dots,K,
\end{split}
\end{align}
and
\begin{align}\label{Benko_comp}
\begin{split}
u(t)&=m_{\!_{0}} + \sum^{\ell}_{r=1} b_{\!_{r}} \sin(2 \pi r f_{\!_{m}}t+ \varphi^b_{\!_{r}}),\\
h_{\!_{1,k}}(t)&=\left [ a_{\!_{k}}+g^A_{\!_{k}}(t) \right ] \sin [ \varphi_{\!_{k}} + g^F_{\!_{k}}(t)], \quad k=1,\dots,K, \\
h_{\!_{2,k}}(t)&=\left [ a_{\!_{k}}+g^A_{\!_{k}}(t) \right ] \cos[ \varphi_{\!_{k}} + g^F_{\!_{k}}(t)], \quad k=1,\dots,K. 
\end{split}
\end{align}
We now show how equations \eqref{Benko_comp2011} and \eqref{Benko_comp} allow to compare our model in equation \eqref{mod_model} with the models proposed by \cite{Benko2011} and \cite{Benko2018}, respectively. Comparing the models in equations \eqref{mod_model} and \eqref{Benko_model0}, time-varying trend and amplitudes of the model in equation \eqref{mod_model} are expressed as
\begin{align}\label{comp_m_u0}
\begin{split}
m(t)&=v(t),\\
g_{\!_{\ell,k}}(t)&=w_{\!_{\ell,k}}(t), \quad  \ell=1,2, \quad k=1,\dots,K.
\end{split}
\end{align}
At the same time, comparing the model in equation \eqref{mod_model} with the model in equation \eqref{Benko_model}, time-varying trend and amplitudes of the model in equation \eqref{mod_model} are
\begin{align}\label{comp_m_u}
\begin{split}
m(t)&=u(t),\\
g_{\!_{\ell,k}}(t)&=h_{\!_{\ell,k}}(t), \quad  \ell=1,2, \quad k=1,\dots,K,
\end{split}
\end{align}
the ordinary frequency being $f_{\!_{k}}=k f_{\!_{0}}$.

As we can see in equations \eqref{comp_m_u0} and \eqref{comp_m_u}, the functions $m(t)$ and $g_{\!_{\ell,k}}(t)$ incorporate the amplitude and frequency modulation functions $g^M(t)$ and $g^M_{\!_{k}}(t)$ in equations \eqref{gMt1} and \eqref{gMt2}. In this sense, the limitation of our approach is that it does not aim at identifying the amplitude and frequency modulating functions $g^M(t)$ and $g^M_{\!_{k}}(t)$ in equations \eqref{gMt1} and \eqref{gMt2}. On the other hand, the benefit of our approach from the estimation viewpoint is twofold. An important advantage of our model in equation \eqref{mod_model} over the models in equations \eqref{Benko_model0} and \eqref{Benko_model} is that, the modulating frequency $f_{\!_{m}}$ does not need to be estimated. In other words, in order to describe statistically a Blazhko light curve using our model in equation \eqref{mod_model}, we only need to estimate $f_{\!_{0}}$. If the observed time series is indeed a Blazhko light curve, 
the modulating frequency $f_{\!_{m}}$ is included in the non-parametric trend $m(t)$ and amplitude $g_{\!_{\ell,k}}$ of our model in equation \eqref{mod_model}. Moreover, assuming that the frequencies are known, for our model in equation \eqref{mod_model} we only need to estimate the functions $m(\cdot)$ and $g_{\!_{\ell,k}}(\cdot)$, whereas for the model in equations \eqref{Benko_model0} and \eqref{Benko_model} the estimated parameters are the amplitudes 
$a^A_{\!_{0}}$, $a_{\!_{0}}$, $m_{\!_{0}}$, $a_{\!_{k}}$'s, $b_{\!_{r}}$'s, and $a_{\!_{kj}}^M$'s, the phases $\varphi^b_{\!_{r}}$'s, $\varphi_{\!_{k}}$'s, $\varphi^M_{\!_{j}}$'s, and $\varphi^M_{\!_{kj}}$'s.  

\section{ESTIMATION}\label{Sec:estimation}
In Section~\ref{sec:PLS} we define estimators of the unknown trend $m(\cdot)$, amplitudes $\{g_{\!_{\ell,k}}(\cdot), \, \ell=1,2,\, k=1,\dots,K\}$, and variance $\sigma^2_{\!{z}}$ of the model in equation \eqref{mod_model}, and in Section~\ref{CV} we explain how to select the tuning parameters associated to the $B$-splines and the penalization used in the estimation method. We denote by $N$ the sample size, $T=t_N-t_1$ the time span, $J$ the number of $B$-splines basis, $d$ the degree of the $B$-splines, $K$ the number of harmonics components, $r$ the order of the penalty, and $M$ the number of replications in Monte Carlo simulations.

We performed our calculations using the R Language for
Statistical Computing \citep{R2021}. Our codes combine existing functions (available as part of {\tt R} packages) with our own development. The computations implemented in this paper are available as a GitHub public code repository\footnote{  \url{https://github.com/DarlinSoto/Modulation-models}.}.

\subsection{Penalized Least squares}\label{sec:PLS}
As mentioned in Section~\ref{Sec:model}, we use $B$-splines to estimate the trend and amplitudes of model given by equation \eqref{mod_model}. The smooth trend function $m(t_i)$ is modeled as a linear combination of $B$-splines basis
\begin{equation*}
m(t_i) = \sum^J_{j=1} \alpha_j B_j(t_i), \quad i=1,\dots,N,
\end{equation*}
which can be written in matrix notation as
$$\vect{m} = \matr{B}\vect{\alpha},$$
where $\vect{m}=[m(t_1),\dots,m(t_N)]\tp$, $\matr{B}=[B_{ij}]=[B_j(t_i)]$ is the $N \times J$ basis matrix ($i=1,\dots,N$, $j=1,\dots,J$) and $\vect{\alpha}=(\alpha_{\!_{1}},\dots,\alpha_{\!_{J}})\tp$. The exact definition of $B$-splines is given in Appendix \ref{Ape_B_splines}. 

The smooth amplitude functions, $g_{\!_{\ell,k}}(t_i)$, $\ell=1,2$, are modeled in the same way:
\begin{equation*}
g_{\!_{1,k}}(t_i) = \sum^J_{j=1} \beta_{\!_{k,j}} B_{\!_{j}}(t_i), \qquad g_{\!_{2,k}}(t_i) = \sum^J_{j=1} \gamma_{\!_{k,j}} B_{\!_{j}}(t_i), \qquad k=1,\dots,K.
\end{equation*}
In matrix notation
$$\vect{g}_{\!_{1,k}} = \matr{B} \vect{\beta}_{\!_{k}} \text{ and } \vect{g}_{\!_{2,k}} = \matr{B} \vect{\gamma}_{\!_{k}}, \qquad k=1,\dots,K,$$
where $\vect{\beta}_{\!_{k}}=(\beta_{\!_{k,1}},\dots,\beta_{\!_{k,J}})\tp$, $\vect{\gamma}_{\!_{k}}=(\gamma_{\!_{k,1}},\dots,\gamma_{\!_{k,J}})\tp$, and $\vect{g}_{\!_{\ell,k}}=[g_{\!_{\ell,k}}(t_1),\dots,g_{\!_{\ell,k}}(t_N)]\tp,$ $\ell=1,2$, $k=1,\dots,K$. Thus, $\vect{\alpha}$, $\vect{\beta}_{\!_{k}}$, and $\vect{\gamma}_{\!_{k}}$, $k=1,\dots,K$, are vectors associated to the trend and amplitudes, respectively. We define the $N \times N$ matrices $\matr{C}_{\!_{k}}$ and $\matr{S}_{\!_{k}}$ as
$$\matr{C}_{\!_{k}}=\diag \{ \cos(w_{\!_{k}}t_1),\dots,\cos(w_{\!_{k}}t_N) \} \text{ and } \matr{S}_{\!_{k}}=\diag \{ \sin(w_{\!_{k}}t_1),\dots,\sin(w_{\!_{k}}t_N) \}, \qquad k=1,\dots,K.$$ 

Thus, the model for the expected value of $\vect{Y}$, in matrix notation, can be expressed as
\begin{equation*}
\mean{\vect{Y}}=\vect{\mu}=\matr{\mathcal{B}} \vect{\theta},
\end{equation*}
where $\matr{\mathcal{B}}$ is the $N \times c$ design matrix given by
\begin{equation*}
\matr{\mathcal{B}}= [\matr{B}|\matr{C}_{\!_{1}} \matr{B}|\dots|\matr{C}_{\!_{K}} \matr{B}|\matr{S}_{\!_{1}} \matr{B} | \dots|\matr{S}_{\!_{K}} \matr{B}],
\end{equation*}
with $c=J(2K+1)$, and 
\begin{equation*}
\vect{\theta}=(\vect{\alpha}\tp,\vect{\beta}_{\!_{1}}\tp,\dots,\vect{\beta}_{\!_{K}}\tp,\vect{\gamma}_{\!_{1}}\tp,\dots,\vect{\gamma}_{\!_{K}}\tp)\tp
\end{equation*}
is the vector of regression coefficients of length $c$. 

The \textit{ordinary least squares} (OLS) estimator of $\vect{\theta}$ is the vector $\widehat{\vect{\theta}}_{\!_{\text{OLS}}}$ which minimizes the sum of squares
\begin{equation*}
M_{\vect{\theta}}=|| \vect{Y} - \matr{\mathcal{B}} \vect{\theta} ||^2.
\end{equation*} 
Equating to zero the partial derivatives with respect to each component of $\vect{\theta}$ and assuming (as we shall) that $\matr{\mathcal{B}}\tp \matr{\mathcal{B}}$ is nonsingular, the estimator of $\vect{\theta}$ is
\begin{equation*}
\widehat{\vect{\theta}}_{\!_{\text{OLS}}}=( \matr{\mathcal{B}}\tp \matr{\mathcal{B}})^{-1} \matr{\mathcal{B}}\tp \vect{Y}.
\end{equation*}
The OLS estimate also maximizes the likelihood of the observations when the errors $z_{\!_{1}},\dots,z_{\!_{N}}$ are independent and identically distributed (iid) and Gaussian.

The size of the basis determines the amount of smoothing of the fitted curves. The larger the value of $J$, the bumpier the fitting will be. To avoid overfitting, \cite{Eilers1996}  proposed a penalty on the (high-order) finite differences of the coefficients
\begin{equation*}
M^*_{\vect{\theta}}=|| \vect{Y} - \matr{\mathcal{B}} \vect{\theta}||^2 + \tau_{\!_{1}} || \matr{D}_{\!{r}} \vect{\alpha} ||^2 + \sum^{K}_{k=1} \left \{ \tau_{\!_{2k}} || \matr{D}_{\!{r}} \vect{\beta}_{\!_{k}} ||^2 + \tau_{\!_{2k+1}} || \matr{D}_{\!{r}} \vect{\gamma}_{\!_{k}} ||^2\right \},
\end{equation*}
where $\{\tau_{\!_{k}}$, $k=1,\dots,2K+1\}$ are positive regularization parameters that control the smoothness of the curve, penalizing the coefficients that are far apart from one another. If $\tau_{\!_{k}}=0$, $k=1,\dots,2K+1$,  we have the standard normal equations of linear regression with a $B$-splines basis. The larger the value of $\tau_{\!_{k}}$, the closer the coefficient $\vect{\theta}$ is to zero. When $\tau_{\!_{k}} \rightarrow \infty$  we obtain a polynomial fit. The matrix $\matr{D}_{\!{r}}$ constructs $r$th order differences of a vector $\vect{\eta}$ as
\begin{equation*}
\matr{D}_{\!{r}} \vect{\eta} = \Delta^{r} \vect{\eta}.
\end{equation*}
The first difference of $\vect{\eta}$, $\Delta^1 \vect{\eta}$,  is the vector with elements $\vect{\eta}_{\!_{l+1}}-\vect{\eta}_{\!_{l}}$. Repeated  differencing applied to $\Delta \vect{\eta}$ results in higher differences, such as  $\Delta^2 \vect{\eta}$ and $\Delta^3 \vect{\eta}$.

The penalties can be represented as $\vect{\theta}\tp \matr{P} \vect{\theta}$ with the block-diagonal matrix $\matr{P}=\matr{T} \otimes \matr{D}_{\!{r}}\tp \matr{D}_{\!{r}}$ and $\matr{T}=\diag \{ \tau_{\!_{1}},\tau_{\!_{2}},\tau_{\!_{3}},\dots,\tau_{\!_{2K+1}} \}$. Then, minimizing 
\begin{equation*}
M^*_{\vect{\theta}}=|| \vect{Y} - \matr{\mathcal{B}} \vect{\theta}||^2 + \vect{\theta}\tp \matr{P} \vect{\theta}
\end{equation*}
with respect to $\vect{\theta}$, the \textit{penalized ordinary least squares estimator} (POLS) of $\vect{\theta}$ is
\begin{equation}\label{theta_hat}
\widehat{\vect{\theta}}_{\!_{\text{POLS}}}=( \matr{\mathcal{B}}\tp \matr{\mathcal{B}}+\matr{P})^{-1} \matr{\mathcal{B}}\tp \vect{Y}.
\end{equation}
The prediction of $Y$ at time $t_i$ is given by
\begin{equation}\label{pred_POLS}
\widehat{Y}_{\!_{i}}= \widehat{\mu}(t_i)=\vect{\mathcal{B}}(t_i)\tp \widehat{\vect{\theta}}_{\!_{\text{POLS}}},
\end{equation}
where $\vect{\mathcal{B}}(t_i)$ is the $i$th row of $\matr{\mathcal{B}}$, the residuals are $\widehat{\vect{z}}=\vect{Y}-\widehat{\vect{Y}}$, with $\widehat{\vect{Y}}=(\widehat{Y}_{\!_{1}},\dots,\widehat{Y}_{\!_{N}})\tp$, and the mean square error (MSE) is $\text{MSE}=N^{-1} \sum^N_{i=1} (Y_{\!_{i}}-\widehat{Y}_{\!_{i}})^2$.

The estimators of the trend $\vect{m}$ and amplitudes $\{\vect{g}_{\!_{\ell,k}},\, \ell=1,2,\, k=1,\dots,K\}$, are
\begin{align}\label{m_g_hat}
\widehat{\vect{m}} = \matr{B}\widehat{\vect{\alpha}}, \quad \widehat{\vect{g}}_{\!_{1,k}} = \matr{B}\widehat{\vect{\beta}}_{\!_{k}}, \quad \widehat{\vect{g}}_{\!_{2,k}} = \matr{B}\widehat{\vect{\gamma}}_{\!_{k}}.
\end{align}

Another parameter of interest is the variance of the errors, $\sigma^2_{\!{z}}$,  which can be estimated by
\begin{equation*}
\widehat{\sigma}^2_{\!{z}}=[N-\trace(\widehat{\matr{S}})]^{-1} \sum^N_{i=1} \left \{ Y_{\!_{i}} - \vect{\mathcal{B}}(t_i)\tp \widehat{\vect{\theta}}_{\!_{\text{POLS}}} \right \}^2
\end{equation*}
where $\widehat{\matr{S}}=\matr{\mathcal{B}} ( \matr{\mathcal{B}}\tp \matr{\mathcal{B}}+\matr{P})^{-1} \matr{\mathcal{B}}\tp.$

In addition to the point estimate, interval estimation for $\widehat{Y}_{\!_{i}}$ is often of interest and is easy to construct. In Appendix~\ref{app_CI} we derive parametric and non-parametric confidence intervals for $\widehat{Y}_{\!_{i}}$. 

\subsection{Automatic selection of the tunable parameters}\label{CV}
Before calculating the estimator in equation \eqref{theta_hat}, it is necessary to select the tuning parameters $\vect{\tau}=(\tau_{\!_{1}},\tau_{\!_{2}},\dots,\tau_{\!_{2K+1}})\tp$. To choose the tuning parameters, we propose to use the Akaike information criterion (AIC).

The AIC penalizes the log-likelihood of a fitted model by considering the effective number of parameters. The definition of AIC given by \cite{Hastie2004} is
$$\text{AIC}(\vect{\tau})=\overline{\text{err}}(\vect{\tau}) + 2\frac{\df}{N} \widehat{\sigma}^2_{\!{0}},$$
where $\overline{\text{err}}(\vect{\tau})$ corresponds to the mean square error in the case of Gaussian errors, $\df$ is the effective number of parameters, $N$ the number of observations used to fit the model, and $\widehat{\sigma}^2_{\!{0}}$ is given by the variance of the residuals from the $\widehat{Y}_{\!_{i}}$ that are computed when $\vect{\tau}=\vect{0}_{\!_{2K+1}}$.

The value for $\vect{\tau}$ is chosen by minimizing the AIC, which is computed as
\begin{equation}\label{AIC}
\text{AIC}(\vect{\tau})=\frac{1}{N}\sum^{N}_{i=1} \left \{ Y_{\!_{i}} - \vect{\mathcal{B}}(t_i)\tp \widehat{\vect{\theta}}_{\!_{\text{POLS}}} \right \}^2 + 2\frac{ \trace(\widehat{\matr{S}}) }{N} \widehat{\sigma}^2_{\!{0}},
\end{equation} 

The AIC given by equation \eqref{AIC} can also be used to select the number of $B$-splines $J$, the degree $d$ of $B$-spline, the order of penalty $r$, and the number of harmonic components $K$.

\begin{figure}[htb!]
\centering
\includegraphics[width=18cm]{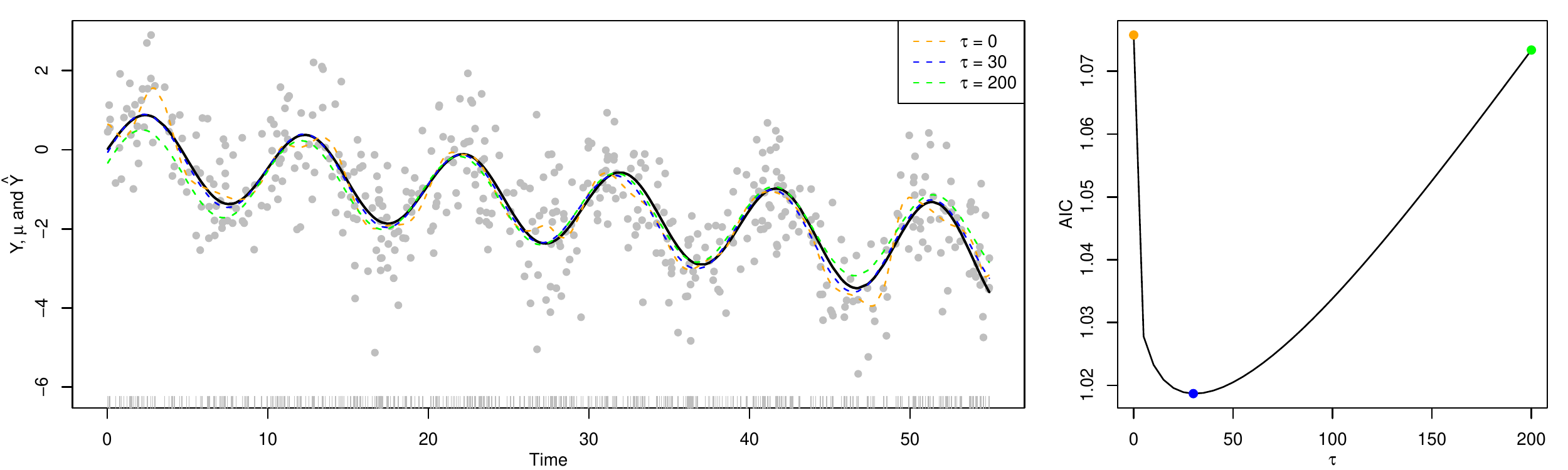}
\caption{Automatic selection of the tunable parameters presented in Section~\ref{CV}. Left: Data (grey dots) simulated according to the model defined by equation \eqref{mod_model}, with $N=500$, $\mu(t_i)=-0.05t_i-(-0.0002t_i+0.0003t_i^2)\cos(0.2 \pi t_i) + (1-0.0005t_i) \sin(0.2 \pi t_i)$ (black curve), and where the errors follow a Gaussian distribution with zero mean and variance $\sigma^2_{\!{z}}=1$. Time is unequally spaced obtained from a uniform distribution $U(0,55)$ (grey ticks on the horizontal axis). We illustrate three estimates of $\vect{Y}$ corresponding to three different specifications of $\tau_{\!{j}}$, with $j=1,2,3$: $\tau_{\!{j}}=0$ (orange curve), $\tau_{\!{j}}=30$ (blue curve), and $\tau_{\!{j}}=200$ (green curve). Right: Values of the AIC in equation \eqref{AIC}, obtained from the simulated and estimated light curve, corresponding to forty-one equally spaced values of $\tau$ ranging from 0 to 200. The three points (orange, blue and green) on the AIC curve correspond to the three fits presented in the left-hand plot of the figure.}\label{Figure:PSE_WN}
\end{figure}

In Figure~\ref{Figure:PSE_WN}, we have generated $N=500$ observations from the model described in equation \eqref{mod_model}, with the Gaussian errors $\{z_i,\, 1\le i\le 500\}$ being simulated using the {\tt R} function {\tt rnorm}. We consider the following artificial signal:  $$\mu(t_i)=-0.05t_i-(-0.0002t_i+0.0003t_i^2)\cos(0.2 \pi t_i) + (1-0.0005t_i) \sin(0.2 \pi t_i),$$ 
with the errors following a Gaussian distribution with zero mean and variance $\sigma^2_{\!{z}}=1$. Time $t$ is unequally spaced, and was obtained from a uniform distribution $U(\theta_1,\theta_2)$ with $\theta_1=0$ and $\theta_2=55$ using the {\tt R} function {\tt runif}. In the first plot of Figure~\ref{Figure:PSE_WN} the observations $\vect{Y}$ are represented by the grey points, and the mean $\vect{\mu}$ by the black curve. The orange, blue and green curves illustrate three possible estimates for $\vect{Y}$ obtained using the method described in the Section \ref{sec:PLS} with increasing smoothing parameters. The orange line is the fit obtained with $\tau_{\!_{j}}=0$, $j=1,2,3$: the corresponding $\widehat{\vect{Y}}$ matches the data well, but fits the true $\vect{\mu}$ poorly because it is wiggly. The blue curve is obtained using the smoothing parameters $\tau_{\!{j}}=30,\,j=1,2,3$, and the green curve is obtained using $\tau_{\!{j}}=200,\,j=1,2,3$. In the second plot of Figure~\ref{Figure:PSE_WN}, we observe that the optimal tuning parameters are $\tau_{\!{j}}=30,\,j=1,2,3$, and as the values of $\vect{\tau}$ increase the obtained curve fits the observed data less closely.

\section{DETECTING SERIAL CORRELATION}\label{Sec:corr}

A statistical model is an approximation to the true process that generates the observed data. After fitting the model given by equation \eqref{mod_model}, it is necessary to check whether the residuals obtained from the fit behave like a white noise process. A significant departure from this assumption suggests the inadequacy of the assumed form of the model. Thus, it is important to assess whether the residuals follow a white noise process.

Detecting serial correlation becomes more challenging when the available observations are unequally spaced in time. If the observations are unequally spaced, so are the errors. In order to study the spectral density of the residuals obtained from the fitted model, in this section we derive the mathematical definition of the spectrum for unequally spaced time series. 

Before presenting our approach, we briefly review the results given by \cite{Deeming1975} about the relationships between the periodogram, the spectral density (PSD), and the autocorrelation function for {\it continuous} time series. Then, we extend the results given by \cite{Deeming1975} to the case of {\it discrete} time series.

Let $\{\varepsilon_{\!_{i}}\}$ be a {\it continuous}, 
zero-mean stationary times series with spectral density \begin{equation*}
P_{\!{\varepsilon}}(\lambda)=\int^{\infty}_{- \infty} r_{\!{\varepsilon}}(h) \exp(i  \lambda h) dh, \quad -\infty < \lambda < \infty,
\end{equation*}
and autocovariance function
$$r_{\!{\varepsilon}}(h)=\int^{\infty}_{- \infty} P_{\!{\varepsilon}}(\lambda) \exp(- i  \lambda h) d\lambda, \quad h \in \mathbb{R}.$$

Consider a time series $\varepsilon_{\!_{1}},\dots,\varepsilon_{\!_{N}}$ with spectrum $P_{\!{\varepsilon}}(\cdot)$ and autocovariance function
$r_{\!{\varepsilon}}(\cdot)$, and assume that the observations $\varepsilon_{\!_{1}},\dots,\varepsilon_{\!_{N}}$ are  obtained at {\it unequally spaced} times $t_{\!_1},\dots, t_{\!_N}$, respectively. The periodogram of $\vect{\varepsilon}=(\varepsilon_{\!_{1}},\dots,\varepsilon_{\!_{N}})\tp$ at frequency $\lambda$ is defined as
\begin{equation}\label{periodogram}
I_{\,\!{\varepsilon}}(\lambda)=\sum^N_{k=1} \sum^N_{j=1} \varepsilon_{\!_{k}} \varepsilon_{\!_{j}} \exp(i  \lambda [t_k - t_j]), \quad \lambda=2\pi f.
\end{equation}
\cite{Deeming1975} proved that the expectation of the periodogram of $\{\varepsilon_{\!_{i}}\}$ in equation \eqref{periodogram} is
\begin{equation}\label{Deeming_result}
\mean{I_{\,\!{\varepsilon}}(\lambda)}=P_{\!{\varepsilon}}(\lambda) \star W_{\!{\varepsilon}}(\lambda),
\end{equation}
where
$W_{\!{\varepsilon}}(\lambda)$ is the power spectral window given by
\begin{equation*}
W_{\!{\varepsilon}}(\lambda)=\sum^N_{j=1} \sum^N_{k=1} \exp(i  \lambda  [t_k - t_j]),
\end{equation*}
and $P_{\!{\varepsilon}}(\lambda) \star W_{\!{\varepsilon}}(\lambda)$ is the continuous convolution of $P_{\!{\varepsilon}}(\lambda)$ with $W_{\!{\varepsilon}}(\lambda)$ defined as
$$P_{\!{\varepsilon}}(\lambda) \star W_{\!{\varepsilon}}(\lambda)=\int^{\infty}_{-\infty} P_{\!{\varepsilon}}(\omega) W_{\!{\varepsilon}}(\lambda-\omega) d\omega.$$

The following lemma states that  it is possible to extend the result in equation \eqref{Deeming_result} to the case of a {\it discrete} zero-mean stationary times series that is generated according to {\it equally spaced} times but observed at {\it unequally spaced} times. The lemma applies to unequally spaced time points $t_i$ with index $i$ belonging to a subset $\mathcal{I}$ of the set $\mathbb{N}=\{ 1,2,\dots\}$ of positive integers. 

\begin{lem}\label{lem:exp}
Let $\{\varepsilon_{\!_{i}}, \text{ } t_i=t_0+i\Delta, \text{ } \Delta>0, \text{ } i\in\mathcal{I}\subseteq \mathbb{N}\}$ be a zero-mean, stationary, discrete time series with spectral density
\begin{equation}\label{spectral_density_disc}
P_{\!{\varepsilon}}(\lambda)=\frac{1}{2\pi} \sum^{\infty}_{h=-\infty} \exp(i \lambda h \Delta ) r_{\!{\varepsilon}}(h), \quad - \infty \leq \lambda \leq \infty ,
\end{equation}
with autocovariance function defined as $r_{\!{\varepsilon}}(h)=\mean{\varepsilon_{\!_{k}} \varepsilon_{\!_{j}}}$, with $k=j+|h|$, $h \in \mathbb{Z}$, that can be expressed in term of the spectral density in equation \eqref{spectral_density_disc} as
\begin{equation}\label{autocovar_disc}
r_{\!{\varepsilon}}(h)= \frac{2\pi}{N_{\,\!\mathcal{I}}} \sum^{N_{\,\!\mathcal{I}}}_{j=1} \exp(-i \lambda_{j} h \Delta) P_{\!{\varepsilon}}(\lambda_{j}), \quad h \in \mathbb{Z},
\end{equation}
where $\lambda_{j}=2\pi f_{\!_{j}}$, with  $f_{\!_{j}}=j/(N_{\mathcal{I}} \Delta)$ and $N_{\mathcal{I}}=\max\mathcal{\{I\}}$. Then, the expectation of the periodogram in equation \eqref{periodogram}  obtained from $\{\varepsilon_{\!_{i}},  i\in\mathcal{I}\}$
is
\begin{equation}\label{exp_perio_disc}
\mean{I_{\,\!{\varepsilon}}(\lambda)}=\frac{2\pi}{N_{\,\!\mathcal{I}}} \,P_{\!{\varepsilon}}(\lambda) * W_{\!{\varepsilon}}(\lambda),
\end{equation}
with power spectral window given by
\begin{equation}\label{power_spectral_window}
W_{\!{\varepsilon}}(\lambda)=\sum_{k \in \mathcal{I}} \sum_{j \in \mathcal{I}} \exp(i  \lambda  [t_k - t_j])
\end{equation}
and $P_{\!{\varepsilon}}(\lambda) * W_{\!{\varepsilon}}(\lambda)$ is the discrete convolution of $P_{\!{\varepsilon}}(\lambda)$ with $ W_{\!{\varepsilon}}(\lambda)$ defined as
$$P_{\!{\varepsilon}}(\lambda) * W_{\!{\varepsilon}}(\lambda)=\sum^{N_{\,\!\mathcal{I}}}_{j=1} P_{\!{\varepsilon}}(\omega_{\!{j}}) * W_{\!{\varepsilon}}(\lambda-\omega_{\!{j}}), \quad \omega_{\!{j}}=2\pi f_{\!_{j}}, \quad f_{\!_{j}}=\frac{j}{N_{\,\!\mathcal{I}} \Delta}.$$
\end{lem}

Our result in equation~\eqref{exp_perio_disc} differs from the result  by \cite{Deeming1975} in equation~\eqref{Deeming_result}.  \cite{Deeming1975} proved that the expectation of both discrete and continuous Fourier transforms of a {\it continuous stochastic process} $f(t)$ \citep[in the sense of equations (31) and (32) in][]{Deeming1975} is equal to the {\it continuous convolution} of the spectral density of $f(t)$ with a spectral window \citep[see equations (33) and (36) in][] {Deeming1975}. In Lemma~\ref{lem:exp}, instead, we prove that the expectation of the discrete Fourier transform of the {\it discrete stochastic process} $\varepsilon_t$ is equal to the {\it discrete convolution} of the spectral density of $\varepsilon_t$ with a spectral window (up to the constant $2\pi/N_\mathcal{I}$). 

When the time series is generated according to an equally spaced stochastic process and the observations are equally spaced, the periodogram is an (asymptotically) unbiased estimator of the spectral density  \citep[see][page 418]{Priestley1981}. However, when the observations are unequally spaced it does not make sense to estimate the spectral density in the same way. This is due to the  \textit{power spectral window} $W_{\!{\varepsilon}}(\lambda)$ in equations \eqref{exp_perio_disc}-\eqref{power_spectral_window}. Nevertheless, as we show in the following proposition, it is possible to disentangle the spectral density $P_{\!{\varepsilon}}(\lambda)$ from the spectral window $W_{\!{\varepsilon}}(\lambda)$.  
\begin{prop}\label{prop:discrete}
Let $\mathcal{F} \{ g_{\!_{j}} \}[k]$ denote the Discrete Fourier Transform of the sequence of $m$ numbers $g_{\!_{1}},\dots,g_{\!_{m}}$ into another sequence $h_{\!_{1}},\dots,h_{\!_{m}}$, that is,
\begin{equation}\label{CFT-text}
h_{\!_{k}}=\mathcal{F} \{ g_{\!_{j}} \} [k]=\sum^m_{j=1} g_{\!_{j}} \exp(-i k j 2\pi/m), \quad k=1,\dots,m.
\end{equation}
Accordingly, define $\mathcal{F}^{-1} \{ h_{\!_{k}} \}[j]$ as the Inverse Discrete Fourier Transform of the sequence $h_{\!_{1}},\dots,h_{\!_{m}}$ into another sequence $g_{\!_{1}},\dots,g_{\!_{m}}$, that is,
\begin{equation}\label{inv_CFT-text}
g_{\!_{j}}=\mathcal{F}^{-1} \{ h_{\!_{k}} \} [j]=\frac{1}{m} \sum^m_{k=1} h_{\!_{k}} \exp(i k j 2\pi/m), \quad j=1,\dots,m.
\end{equation}
 Assume that $\{\varepsilon_{\!_{i}},  i\in\mathcal{I}\}$ satisfy the same conditions as in Lemma~\ref{lem:exp}. Then
 we can write the spectral density $P_{\!{\varepsilon}}(\lambda_{j})$ in equation \eqref{spectral_density_disc} at frequency $\lambda_{j}=2 \pi f_{\!_{j}}$, with $f_{\!_{j}}=j/(N_{\,\!\mathcal{I}}\Delta)$, as \begin{equation} \label{spectral_density_hat-text}
P_{\!{\varepsilon}}(\lambda_{j})=\frac{N_{\,\!\mathcal{I}}}{2\pi} \mathcal{F}^{-1}\left \{ \frac{\mathcal{F} \{ \mean{I_{\!_{\varepsilon}}(\lambda_{j})} \}[k] }{\mathcal{F} \{ W_{\!{\varepsilon}}(\lambda_{j}) \}[k] } \right \} [j],\qquad j=1,\dots,N_{\,\!\mathcal{I}}.
\end{equation}
\end{prop}
The proofs of Lemma~\ref{lem:exp} and Proposition~\ref{prop:discrete} are given in Appendix~\ref{app_proofs}. Equation \eqref{spectral_density_hat-text} suggests that in order to estimate $P_{\!{\varepsilon}}(\lambda)$, we need the value of $\Delta$ and $W_{\!{\varepsilon}}(\lambda)$, as well as the estimate of $\mean{I_{\,\!{\varepsilon}}(\lambda)}$. Notice that, in general, the time series denoted in this section as   $\vect{\varepsilon}=(\varepsilon_{\!_{1}},\dots,\varepsilon_{\!_{N}})\tp$ is possibly autocorrelated, whereas the errors $\vect{z}=(z_{\!_{1}},\dots,z_{\!_{N}})\tp$ of our model in equation \eqref{mod_model} are assumed to be serially uncorrelated. The following algorithm is proposed to establish whether the unequally spaced residuals obtained when fitting our model in equation \eqref{mod_model} are uncorrelated.
\begin{itemize}
\item [i)] Obtain the residuals $\widehat{\vect{z}}=\vect{Y}-\widehat{\vect{Y}}$.
\item [ii)] For each $j=1,\dots,N_{\,\!\mathcal{I}}$, define $\hat{I}_{\,\!{\widehat{z}}}(\lambda_{j})$ as the periodogram in equation \eqref{periodogram} computed upon the residuals $\widehat{\vect{z}}$ obtained in step i), with $\lambda_{j}=\tfrac{2\pi j}{N_{\,\!\mathcal{I}} \Delta}$.
\item [iii)] For each $j=1,\dots,N_{\,\!\mathcal{I}}$, define $W_{\!{\widehat{z}}}(\lambda_{j})$ as the power spectral window in equation \eqref{power_spectral_window} computed upon the residuals $\widehat{\vect{z}}$ obtained in step i), with $\lambda_{j}=\tfrac{2\pi j}{N_{\,\!\mathcal{I}} \Delta}$.
\item [iv)] Smooth the periodogram obtained in step ii) over frequencies, and denote the smoothed periodogram by $\tilde{I}_{\,\!{\widehat{z}}}(\lambda_{j})$. 
\item [v)] Calculate the Discrete Fourier Transform in equation \eqref{CFT-text} of the power spectral window and the periodogram obtained in steps iii) and iv), respectively.
\item [vi)] For each frequency $\lambda_{j}=\tfrac{2\pi j}{N_{\,\!\mathcal{I}} \Delta}$, define the estimated spectral density of the errors $\vect{z}$ as
\begin{equation}\label{PSD_hat}
\widehat{P}_{\!z}(\lambda_{j})=\frac{N_{\,\!\mathcal{I}}}{2\pi} \mathcal{F}^{-1}\left \{ \frac{\mathcal{F} \{ \tilde{I}_{\,\!{\widehat{z}}}(\lambda_{j}) \}[k] }{\mathcal{F} \{ W_{\!{\widehat{z}}}(\lambda_{j}) \}[k] } \right \} [j],\qquad j=1,\dots,N_{\,\!\mathcal{I}},
\end{equation}
where the inverse Fourier transform $\mathcal{F}^{-1}$ is given by equation \eqref{inv_CFT-text}.
\item [vii)] If the estimated spectral density obtained in step vi) does not vary significantly over frequencies, conclude that the errors are uncorrelated over time. 
\end{itemize}

\section{SIMULATION RESULTS}\label{Sec:simres}
In this section we provide Monte Carlo simulations to illustrate the performance of the estimators $\widehat{\mu}(t)$, $\widehat{m}(t)$, $\widehat{g}_{\!_{\ell,k}}(t)$, $\ell=1,2$, $k=1,\dots,K$, defined by equations \eqref{pred_POLS} and \eqref{m_g_hat}, and the estimator of the spectral density in equation~\eqref{spectral_density_hat-text}.

In Section~\ref{sim_tv_model} we simulate unequally spaced observations from the model in equation~\eqref{mod_model}, under two scenarios. 
In the first scenario both trend and amplitudes sinusoidal, whereas in the second scenario trend and amplitudes are polynomial. In Section \ref{sim_blazhko} we simulate a Blazhko light curve and fit the model in equation~\eqref{mod_model}. Finally in Section \ref{sim_deeming} we evaluate the performance of the estimator of the spectral density defined in equation~\eqref{spectral_density_hat-text} of a discrete unequally-spaced time series.

\subsection{Simulating our novel time-varying model}\label{sim_tv_model}
In this section we generate the data according to the model described by equation~\eqref{mod_model} with $K=2$, $N=500$ and time $t$ is unequally spaced obtained form an Uniform distribution $U(\theta_1,\theta_2)$ with $\theta_1=0$ and $\theta_2=1$. In order to illustrate the flexibility of our novel method, we consider two different scenarios for the trend and amplitudes. In the first scenario, we simulate sinusoidal trend and  amplitudes as $m(t)=\sin(2\pi t)$, $g_{\!_{1,1}}(t)=\cos(9 \pi t)$, $g_{\!_{2,1}}(t)=\sin(6 \pi t)$, $g_{\!_{1,2}}(t)=\cos(4 \pi t)$, $g_{\!_{2,2}}(t)=\sin(7\pi t)$, with frequencies $w_{\!_{1}}=40 \pi$, and $w_{\!_{2}}=100 \pi$. In the second scenario, we simulate (global) polynomial trend and amplitudes as  $m(t)=0.2t-5t^2+5.5t^3$, $g_{\!_{1,1}}(t)=4t^3-5t^2$, $g_{\!_{2,1}}(t)=-0.5-0.5t+2.5t^2-0.5t^3$, $g_{\!_{1,2}}(t)=-t+t^2+1.3t^3$, $g_{\!_{2,2}}(t)=0.5+2t^2-3t^3$, with frequencies $w_{\!_{1}}=30 \pi$, and $w_{\!_{2}}=40\pi$. 
  In both scenarios, we assume that the error terms $ \{ z_{\!_{i}}$, $i=1,\dots,N \}$ follow a Gaussian distribution with zero mean and variance $\sigma^2_{\!{z}}=2$. 
  
\begin{figure}[htb!]
\centering
\includegraphics[width=18cm]{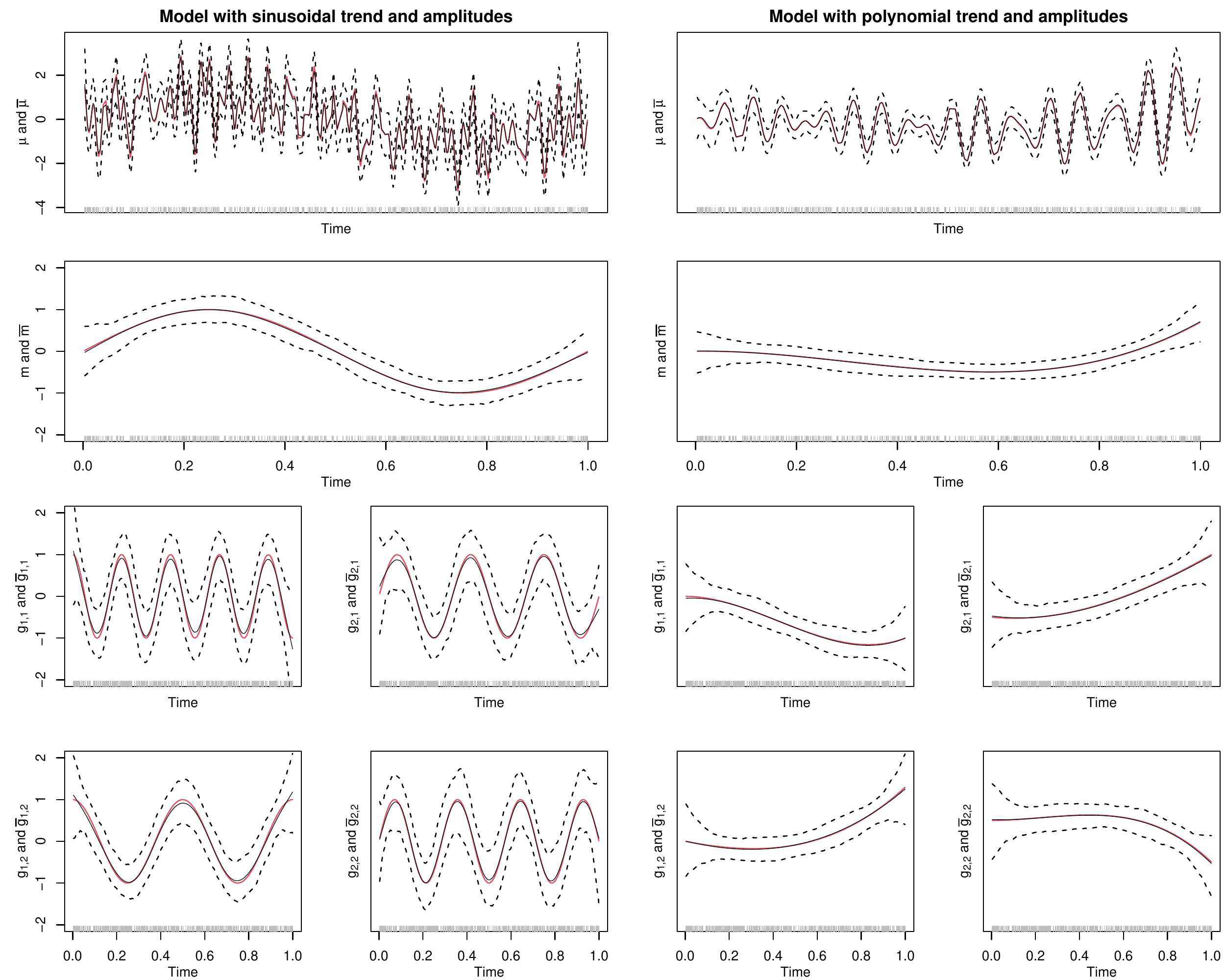}
\caption{Simulation scenarios of Section~\ref{sim_tv_model}: data generated from the model in equation~\eqref{mod_model} with {\it sinusoidal} and {\it polynomial} time-varying trend and amplitudes. Time $t$ is unequally spaced obtained form the Uniform distribution $U(0,1)$. The first column shows the fit of the model in equation~\eqref{mod_model} with sinusoidal trend and amplitudes, whereas the second column shows the fit of the model in equation~\eqref{mod_model} with polynomial trend and amplitudes. From the $M=200$ realizations of our estimators we compute, for each fixed $t$, three averages and confidence intervals. 
The first row shows the true $\mu(t)$ (red solid line), together with the average $\overline{\mu}(t)=\frac{1}{M} \sum^M_{j=1}\widehat{\mu}^{(j)}(t)$ of the estimates $\widehat{\mu}^{(j)}(t)$ (black solid line). The second row shows the true trend $m(t)$ together with the average $\overline{m}(t)=\frac{1}{M} \sum^M_{j=1}\widehat{m}^{(j)}(t)$ of the estimates $\widehat{m}^{(j)}(t)$. The third and fourth rows show the true amplitudes $g_{\!_{\ell,k}}(t)$, $\ell=1,2$, $k=1,2$, together with the average $\overline{g}_{\!_{\ell,k}}(t)=\frac{1}{M} \sum^M_{j=1}\widehat{g}_{\!_{\ell,k}}^{(j)}(t)$ of the estimates $\widehat{g}^{(j)}_{\!_{\ell,k}}(t)$. The non-parametric quantiles (black dashed lines) are the confidence intervals corresponding to the 2.5th and 97.5th order statistics, respectively, see Appendix~\ref{Ape_B_CI_nonpar}.}
\label{Figure:2}
\end{figure}

In both scenarios, we simulate $M=200$ realizations of the model in equation~\eqref{mod_model}. For each $j=1,\dots,M$, we compute the estimate $\widehat{\vect{\theta}}^{(j)}$ defined by equation~\eqref{theta_hat}. In the first scenario, we select the smoothing parameter $\vect{\tau}=(50,1,2,10,1)\tp$, a total number of B-splines $J=33$ of order $d=3$, and an order penalty $r=2$. In the second scenario, we choose the smoothing parameter $\vect{\tau}=(3,3,3,3,3)\tp$, a total number of $B$-splines $J=6$ of order $d=3$, and an order penalty $r=4$. Figure \ref{Figure:2} shows our estimates of $\vect{\mu}$, $\vect{m}$, $\{\vect{g}_{\!_{\ell,k}},\,\ell=1,2,\,k=1,2\}$, and their 95\% confidence intervals. Figure \ref{Figure:2} shows that our model in equation~\eqref{mod_model} fits well the simulated data in both the sinusoidal and polynomial scenarios. That is, the trend and amplitudes are well fitted in both scenarios. The 95\% confidence intervals are constructed in a non-parametric fashion using quantiles; see Appendix \ref{Ape_B_CI_nonpar} for more details.

\subsection{Simulating a Blazhko RR Lyrae light curve characterized by amplitude modulation}\label{sim_blazhko}
We simulate a Blazhko RR Lyrae light curve with amplitude modulation according to \cite{Benko2011} as
\begin{equation}\label{rrl0}
\begin{split}
Y_{\!_{i}}&=\mu(t_i)+z_{\!_{i}}, \quad i=1,\dots,N=1000,\\
\mu(t_i)&=\left [ 1+\frac{U_{\!_{m}}(t_i)}{U_{\!_{c}}} \right ] \,c(t),\\
c(t)&=a_{\!_{0}}+\sum^{4}_{k=1} a_{\!_{k}} \sin(2\pi k f_{\!_{0}} t_i+ \varphi_{\!_{k}}),\\
U_{\!_{m}}(t)&=a_{\!_{m}} \sin(2 \pi f_{\!_{m}} t + \varphi_{\!_{m}}),
\end{split}
\end{equation}
where $c(t)$ is the carrier wave with four harmonic components, $U_{\!_{m}}(t)$ is the modulating signal, $U_{\!_{c}}=a_{\!_{m}}/h$ is the amplitude of the non-modulated light curve, and $\{ z_{\!_{i}}, i=1,\dots,N \}$ are the error terms. The values of the parameters used in equation~\eqref{rrl0} and the time-design are obtained from \cite{Benko2011}. In particular, 
$a_{\!_{m}}=0.1$ mag, $h=1.2$, $a_{\!_{0}}=0.01$ mag, $f_{\!_{m}}=0.05$ d$^{-1}$, $\varphi_{\!_{m}}=270$ degrees, and the values $\{a_{\!_{k}},\, \varphi_{\!_{k}},\,1\leq k\leq 4\}$ are presented in Table 1. We convert the Blazkho phase $\varphi_{\!_{m}}$ and the main phases $\{\varphi_{\!_{k}},\,1\le k\le 4\}$ in equation~\eqref{rrl0} from degrees to radians using the {\tt R} function {\tt NISTdegTOradian} \citep[available in the {\tt R} package {\tt NISTunits2016} by][]{NISTunits2016}. The original time design $\{t_j,\,j=1,\dots,28799\}$ in \cite{Benko2011} is equally spaced. However, variable stars are often observed at irregular intervals. For this reason, in our simulation exercise we sample a subset of the original time points and use this subset to evaluate the performance of our method. We obtain the time design $\{t_i,\,i=1,\dots,1000\}$ in equation~\eqref{rrl0} by sampling the original, equally spaced time design $\{t_j,\,j=1,\dots,28799\}$. We end up with $N=1000$ unequally spaced observations ranging from $t=0.03819$~d to $t=69.37847$~d. The error terms are generated independently from a Gaussian distribution with zero mean and variance $\sigma^2_{\!{z}}=0.005$.

\begin{deluxetable*}{ccll}
\tablenum{1}
\tablecaption{Parameters of simulated RR Lyrae star\label{tab:tablabenko}}
\tablewidth{0pt}
\tablehead{
\colhead{$k$} & \colhead{$kf_{\!_{0}}$} & \colhead{$a_{\!_{k}}$} & \colhead{$\varphi_{\!_{k}}$}\\
 & \colhead{(day$^{-1}$)} & \colhead{(mag)} & \colhead{(degrees)}}
\startdata
1 & 2 & 0.401 & 5.490 \\ 
2 & 4 & 0.171 & 144.040 \\ 
3 & 6 & 0.133 & 285.250 \\ 
4 & 8 & 0.097 & 81.290 \\ 
\enddata
\tablecomments{ Parameters (frequencies, amplitudes, and phases) obtained from \cite{Benko2011}, as explained in Section~\ref{sim_blazhko}}
\end{deluxetable*}

If we consider of our novel model in equation~\eqref{mod_model}, with time-varying trend and amplitudes specified as
\begin{equation}\label{sim_rr}
\begin{split}
m(t_i) &= a_{\!_{0}} \left [ 1+U_{\!_{m}}(t_i)/U_{\!_{c}} \right ], \\ 
g_{\!_{1,k}}(t_i) &= a_{\!_{k}} \cos(\varphi_{\!_{k}})[ 1+U_{\!_{m}}(t_i)/U_{\!_{c}}], \quad k=1,\dots,4, \\
g_{\!_{2,k}}(t_i) &= a_{\!_{k}} \sin(\varphi_{\!_{k}})[ 1+U_{\!_{m}}(t_i)/U_{\!_{c}}], \quad k=1,\dots,4,
\end{split}
\end{equation}
we can rewrite the model in equation~\eqref{rrl0} as a special case of our model given by equation~\eqref{mod_model}. The main advantage of fitting the model in equation~\eqref{mod_model} instead of the model in equation~\eqref{rrl0}, is that one does not need to estimate the parameters $a_{\!_{m}}$, $h$, $a_{\!_{0}}$, $\varphi_{\!_{m}}$, $a_{\!_{k}}$, $\varphi_{\!_{k}}$, $k=1,\dots,4$, $f_{\!_{m}}$.
Moreover, we do not need to adopt any specific functional form for $m(\cdot)$ and $g(\cdot)$, such as those given by equation~\eqref{sim_rr}, because they are well approximated by $B$-splines.

We fit the model in equation~\eqref{mod_model} with $K=4$ to the data generated according to the model in equation~\eqref{rrl0}. We assume that the frequencies $f_{\!_{k}}$, $k=1,\dots,4$, of each harmonic component are known, see Table 1. Also, we use a total of $J=18$ $B$-splines of degree $d=3$, an order penalty $r=1$, and the smoothing parameters $\vect{\tau}=(5,1,0.1,0.1,0.1,0.1,1,0.1,4)\tp$.

\begin{figure}[htb!]
\centering
\includegraphics[width=18cm]{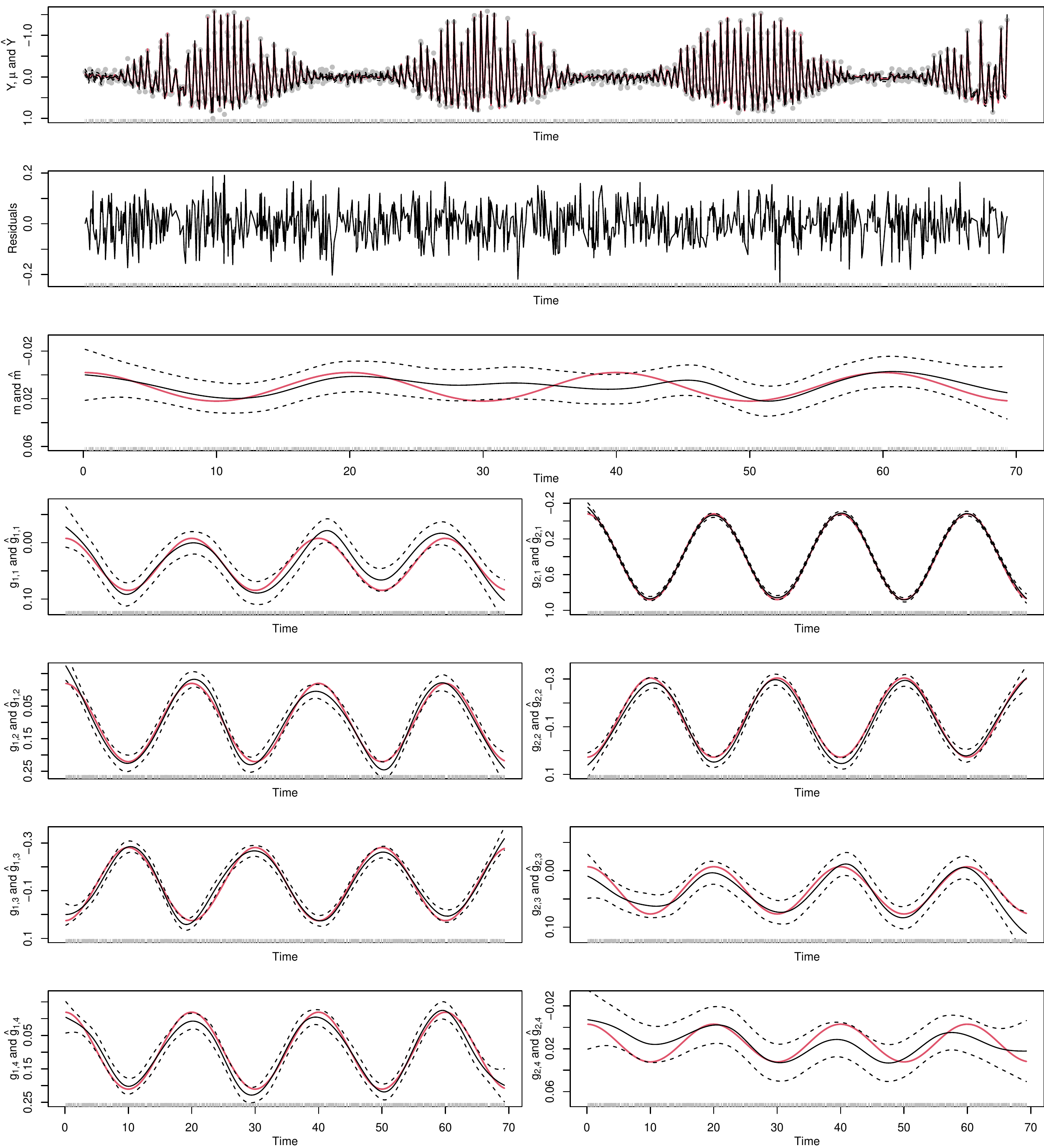}
\caption{Simulated Blazhko RR Lyrae light curve characterized by amplitude modulation, see Section \ref{sim_blazhko}. The first row shows the light curve data of an RR Lyrae star simulated according to the model given by equation~\eqref{rrl0} (grey points), the curve $\mu(t)$ in equation~\eqref{rrl0} (solid-red  line), the prediction $\widehat{Y}_{\!_{i}}$ (solid-black line), and the parametric 95\% confidence intervals (dashed-black lines) obtained according to Appendix~\ref{Ape_B_CI_par}. The second row shows the residuals. The third row shows the time-varying trend ${m}(t_i)$ (red solid lines) simulated according to equation~\eqref{sim_rr}, together with the estimated trend $\widehat{m}(t_i)$ (solid black lines) obtained according to equation  \eqref{m_g_hat}. The last four rows show the  time-varying amplitudes $\{ g_{\!_{\ell,k}}(t_i),\,\, \ell=1,2,\, k=1,\dots,4\}$ (red solid lines) simulated according to equation~\eqref{sim_rr}, together with their estimates $\widehat{g}_{\!_{\ell,k}}(t_i)$ (solid-black lines) given by equation~\eqref{m_g_hat}. The parametric 95\% confidence intervals (dashed-black lines) are  obtained according to Appendix~\ref{Ape_B_CI_par}.}\label{Figure:RRL_1}
\end{figure}

We fit the model in equation~\eqref{mod_model} to the simulated data obtained from the model in equation~\eqref{rrl0}, and present the results in Figure \ref{Figure:RRL_1}. The first row shows the simulation of the amplitude-modulated RR Lyrae light curve given by equation~\eqref{rrl0} (grey points), together with the true and fitted curve (solid-red and solid-black lines, respectively). The second row shows the residuals, and the third and fourth rows show the true trend and amplitudes (red lines) given by equation~\eqref{sim_rr} and their fits (black lines). We observe from Figure \ref{Figure:RRL_1} that the model in equation~\eqref{mod_model} fits well the simulated data. That is, trend and amplitudes are well fitted, and the residuals satisfy the assumption of zero mean and constant variance. The 95\% confidence intervals are constructed in a parametric fashion, see Appendix~\ref{Ape_B_CI_par} for more details.

\subsection{Estimating the spectral density of  unequally spaced time series}\label{sim_deeming}
In this section we estimate the spectral density of unequally spaced time series by means of our novel estimator in equation~\eqref{spectral_density_hat-text}. To this end, we simulate  unequally spaced observations generated from the following AR(2) process: 
\begin{equation}\label{AR2_model}
\begin{split}
\varepsilon_{\!_{i}}&=\phi_{\!_{1}} \varepsilon_{\!_{i-1}}+\phi_{\!_{2}} \varepsilon_{\!_{i-2}}+z_{\!_{i}}\\
z_{\!_{i}}&\sim\mathcal{N}(0,\sigma^2_{\!{z}})\\
t_i&=t_0+i\Delta,
\end{split}
\end{equation}
where $i=1,\dots,N$, with $N=500$ equally spaced observations, starting time $t_0=0.67$, and  $\Delta=0.33$. In order to simulate a realistic AR(2) process, we use the coefficients of the Sunspot Numbers in Example~3.2.9 of \cite{brockwell2016}, where  $\phi_{\!_{1}}=1.318$, $\phi_{\!_{2}}=-0.634$,  and $\sigma^2_{\!{z}}=289.2$. These $\phi$ coefficients ensure the existence of a causal solution 
\begin{equation}\label{causal}
\varepsilon_{\!_{i}}=\sum_{j=0}^\infty \psi_{\!_{j}} z_{\!_{i-j}}
\end{equation} 
 of equation~\eqref{AR2_model}. The time series in equation~\eqref{causal} is {\it causal} in the sense that $\varepsilon$  depends upon current and past (rather than future) values of the error term $z$. We simulate $M=500$ times the AR(2) model given by equation~\eqref{AR2_model} obtaining the observations $\varepsilon^{(m)}_{\!_{1}},\dots,\varepsilon^{(m)}_{\!_{N}}$, $m=1,\dots,M$. Then, in order to obtain unequally spaced observations 
we use the following three steps. 
\begin{enumerate}
\item 
We divide time into 50 blocks, where each block has 10 observations, in a way to preserve the original time series structure. 
\item 
In order to preserve the autocorrelation between the observations, we select randomly 30 blocks and collect the time points corresponding to these blocks, obtaining a new set of time points $\{ t^*_i, i=1,\dots,n \}$, with $n=300$ observations. In contrast to the simulation schemes of Sections~\ref{sim_tv_model} and~\ref{sim_blazhko} where time was sampled randomly, here data sets with uniformly sampled subsets are produced. While the former sampling
is close to the data distribution of large ground-based surveys, the latter is the typical sampling of photometric space telescopes that are dedicated to high-cadence time-series observations, such as  {\em Kepler} \citep[][]{Koch2010}.

\item Finally, we collect the observations $\varepsilon^{(m)}_{\!_{i}}$ corresponding to the new set of time points $\{ t^*_i, i=1,\dots,n \}$ and rename them as $e^{(m)}_{\!_{i}}=\varepsilon^{(m)}_{\!_{i}}$, with $e^{(m)}_{\!_{i}}$ being observed at time $t^*_i$, $i=1,\dots,n $. 
\end{enumerate}
Thus, we obtain the unequally spaced observations $e^{(m)}_{\!_{1}},\dots,e^{(m)}_{\!_{n}}$, which represent a subset of the equally spaced time series $\varepsilon^{(m)}_{\!_{1}},\dots,\varepsilon^{(m)}_{\!_{N}}$. For each $m=1,\dots,M$, and each fixed frequency $\lambda_{j}=2\pi f_{\!_{j}}$, $f=j/(N \Delta)$, $j=1,\dots,N$, we compute the periodogram of $e^{(m)}_{\!_{1}},\dots,e^{(m)}_{\!_{n}}$ as
\begin{equation*}
I^{(m)}_{\,\!{e}}(\lambda_{j})=\sum^n_{k=1} \sum^n_{d=1} e^{(m)}_{\!_{k}} e^{(m)}_{\!_{d}} \exp(i \lambda_{j} [t^*_k - t^*_d]),
\end{equation*}
the average of the periodograms $I^{(m)}_{\,\!{e}}(\lambda_{j})$ as
\begin{equation*}
\overline{I}_{\,\!{e}}(\lambda_{j})=\frac{1}{M} \sum^M_{m=1} I^{(m)}_{\,\!{e}}(\lambda_{j}),
\end{equation*}
and the power spectral window of $e^{(m)}_{\!_{1}},\dots,e^{(m)}_{\!_{n}}$ as
\begin{equation*}
W_{\!{e}}(\lambda_{j})=\sum^{n}_{d=1} \sum^{n}_{k=1} \exp(i \lambda_{j} [t^*_k - t^*_d]).
\end{equation*}

For each frequency $\lambda_{j}$, $j=1,\dots,N$, replacing $\mean{I_{\,\!{e}}(\lambda_{j})}$ with $\overline{I}_{\!{e}}(\lambda_{j})$ and substituting $W_{\!{e}}(\lambda_{j})$ in equation~\eqref{spectral_density_hat-text}, the estimated spectral density of the unequally spaced time series  $e^{(m)}_{\!_{1}},\dots,e^{(m)}_{\!_{n}}$ is given by
\begin{equation}\label{hat_spectral_density_x}
\widehat{P}_{\!{e}}(\lambda_{j})=\frac{N}{2\pi} \mathcal{F}^{-1}\left \{ \frac{\mathcal{F} \{ \overline{I}_{\!{e}}(\lambda_{j}) \}[k] }{\mathcal{F} \{ W_{\!{e}}(\lambda_{j}) \}[k] } \right \} [j],
\end{equation}
A smooth version of the estimated spectral density in equation~\eqref{hat_spectral_density_x} is
\begin{equation}\label{tilde_spectral_density_x}
\widetilde P_{\!{e}}(\lambda_{j}) = \tfrac 1 {N_{\!_{\lambda}}}\sum_{i=1}^{N_{\!_{\lambda}}} K_h(\lambda_{j}-\lambda_i) \widehat{P}_{\!{e}}(\lambda_{j}).
\end{equation}
The rescaled kernel function is defined as $K_h(x)=\tfrac 1 h K(x/h)$, where $K$ is a second order kernel and $h$ is the bandwidth. For this application, we used the Gaussian kernel $K(y)=\tfrac 1{\sqrt{2\pi}}\exp(-y^2/2)$ and a bandwidth $h=0.3$. 

\begin{figure}[h!]
\begin{center}
\includegraphics[width=17cm]{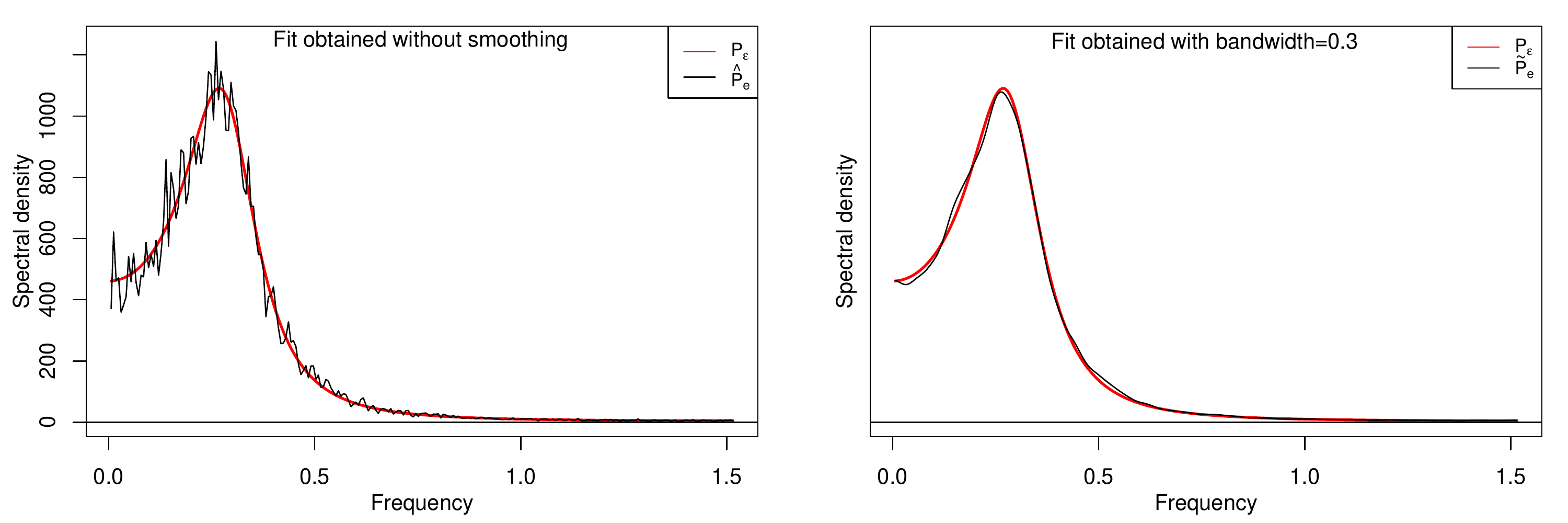}
\end{center}
\caption{Estimated spectral density of the unequally spaced time series sampled by blocks in Section \ref{sim_deeming}. Left: comparison between the true spectral density in equation~\eqref{PSD_AR2} (red line) and the estimated spectral density $\widehat P_{\!_{e}}(\lambda_{j})$ in equation~\eqref{hat_spectral_density_x} (black line) for $j=1\dots,N/2$. Right: comparison between the true spectral density in equation~\eqref{PSD_AR2} (red line) and the smooth estimated spectral density $\widetilde P_{\!_{e}}(\lambda_{j})$ in equation~\eqref{tilde_spectral_density_x} (black line) for $j=1\dots,N/2$. The true spectral density corresponds to the equally spaced time series $\varepsilon_{\!_{i}}$ which follows the AR(2) process given by equation~\eqref{AR2_model}, whereas the estimated spectral density is computed from the unequally spaced observations $e^{(m)}_{\!_{1}},\dots,e^{(m)}_{\!_{n}}$, $m=1,\dots,M$. The unequally spaced time series $e^{(m)}_{\!_{1}},\dots,e^{(m)}_{\!_{n}}$ is obtained as a subset of the time series $ \varepsilon^{(m)}_{\!_{1}},\dots,\varepsilon^{(m)}_{\!_{N}}$. In this example $N=500$, $n=300$, and $M=500$.}\label{SD_X_Y_1}
\end{figure}

Figure \ref{SD_X_Y_1} compares the underlying spectral density $P_{\!{\varepsilon}}(\lambda_{j})$ of the equally spaced time series $\{ \varepsilon_{\!_{i}} \}$, with the estimated spectral densities $\widehat P_{\!{e}}(\lambda_{j})$ and  $\widetilde P_{\!{e}}(\lambda_{j})$  of the unequally spaced time series $\{ e_{\!_{i}} \}$.
The underlying spectral density of the equally spaced time series $\{ \varepsilon_{\!_{i}} \}$, generated by the AR(2) process in equation~\eqref{AR2_model}, is given by
\begin{equation}\label{PSD_AR2}
P_{\!{\varepsilon}}(\lambda_{j})=\frac{\sigma^2_{\!{z}}}{2\pi} \left [ 1 + \phi^2_{\!_{1}} + \phi^2_{\!_{2}} + 2 \phi_{\!_{2}} + 2(\phi_{\!_{1}} \phi_{\!_{2}}-\phi_{\!_{1}}) \cos(\lambda_{j} \Delta) - 4 \phi_{\!_{2}} \cos^2(\lambda_{j} \Delta) \right ]^{-1}.
\end{equation}
The estimated spectral density $\widehat{P}_{\!{e}}(\lambda_{j})$ of the unequally time series $\{ e_{\!_{i}} \}$ is given in equation~\eqref{hat_spectral_density_x}, and its smooth version $\tilde P_{\!{e}}(\lambda_{j})$ in equation~\eqref{tilde_spectral_density_x}. Figure \ref{SD_X_Y_1} shows  that the estimated spectral density of the unequally time series, $\widetilde{P}_{\!{e}}(\lambda_{j})$, fits very well the true spectral density $P_{\!{\varepsilon}}(\lambda_{j})$.

\section{APPLICATION TO REAL DATA} \label{Sec:appl}
In this section, we fit our model in equation~\eqref{mod_model} and the model proposed by \cite{Benko2018} in equation~\eqref{Benko_model} to the same light curve: the V783 Cyg, KIC 5559631. This time series has 61,351 unequally spaced observations, and is available online from the Konkoly Observatory of the Hungarian Academy of Sciences webpage.\footnote{\url{https://konkoly.hu/KIK/data\_en.html}} We choose this particular light curve for two reasons. Firstly, the Blazhko effect of the V783 Cyg time series is known to be characterized by a sinusoidal amplitude and frequency modulation \citep{Benko2014}. The light curve V783 Cyg can be described by $K=15$ significant harmonics with a sinusoidal amplitude and frequency modulations \citep{Benko2014}, which makes V783 Cyg an ideal target for comparing the fits obtained with the models in equations \eqref{mod_model} and \eqref{Benko_model}.  Secondly, these two modulations are well captured and fitted by our novel model in equation~\eqref{mod_model}. 

In order to reduce the computational time, and to satisfy the condition $t_i=t_0+i\Delta$ with $\Delta>0$ and $i\in\mathcal{I}\subseteq \mathbb{N}$ (which is required by Proposition~\ref{prop:discrete}), we analyze a $\approx 78$ d segment of this light curve from $t=827.44$ d to $t=904.9$ d. For this segment, the time-origin and the time-spacing take the values $t_0 = 827.42$~d and $\Delta=0.0204345$~d, respectively, with a total of $N=2101$ unequally spaced observations.

When fitting the models in equations \eqref{mod_model} and \eqref{Benko_model}, the main pulsation and modulation frequencies are not estimated: they take the values $f_{\!_{0}}=1.611084$~d$^{-1}$ and $f_{\!_{m}}=0.036058$~d$^{-1}$ \citep[see][]{Benko2014}, respectively.  Additionally to the $K=15$ significant harmonics fitted by \cite{Benko2014}, we found, after pre-whitening and fitting our model in equation~\eqref{mod_model}, four significant frequencies taking the values 
$f'_{\!_{11}}=18.3254$ d$^{-1}$, $f'_{\!_{12}}=19.9365$ d$^{-1}$, $f'_{\!_{13}}=21.5476$ d$^{-1}$, and  $f'_{\!_{14}}=23.1587$ d$^{-1}$. 
 The values we obtain for $\{f'_j,\,11\le j\le 14\}$ demonstrate that these frequencies are not harmonics of the form $k f_0$, which might suggest that these four are independent frequencies. 
Interestingly, however, we find that the latter belong to a set of fourteen ``reflection frequencies" of the form $\{f'_j=2 f_{\!_N} - (30-j) f_0,\,1\le j\le 14\}$, where $f_{\!_N}=24.46$~d$^{-1}$ is the Nyquist frequency. Among these fourteen frequencies, only the last six $\{f'_j,\,9\le j\le 14\}$ exhibit significant peaks in the Lomb-Scargle periodogram  \citep[computed according to][]{Lomb1976}. However, to avoid over-fitting, we only consider the four frequencies $\{f'_j,\,11\le j\le 14\}$ corresponding to last four peaks of the estimated power spectrum (see the last row of Figure~\ref{Figure:V783_fit}, bottom-right panel). In summary, the only truly independent frequencies are  $f_0$, $f_m$, and $f_{\!_N}$, the other frequencies $\{f_k=k f_0,\,k=1,\dots,15\}$ and $\{f'_j=2 f_{\!_N} - (30-j) f_0,\,1\le j\le 14\}$ being linear combinations (or harmonics) of those.

The frequencies $f'_j$ do not depend on the Blazkho frequency $f_m$, as the information regarding the Blazhko effect is captured by the time-varying trend $m(\cdot)$ and amplitudes $\{g_{\!_{\ell, k}}(\cdot),\,\ell=1,2,\,k=1,\dots,K\}$, see equations \eqref{comp_m_u0}-\eqref{comp_m_u} and Figure~\ref{Figure:V783_fit_g}. We use a different notation ($f'$ rather than $f$) to avoid confusion, since in this case $f_{\!_N}< 16  f_0= 25.77$ d$^{-1}$, and the four frequencies we are considering take value $<24$ d$^{-1}$.

After fitting the models in equations \eqref{mod_model} and \eqref{Benko_model}, we compute their residuals and estimate their spectral densities using equation~\eqref{PSD_hat}. To estimate the spectral densities according to the procedure in Section \ref{Sec:corr}, we adopt the Gaussian kernel $K(y)=\tfrac 1{\sqrt{2\pi}}\exp(-y^2/2)$ with a bandwidth $h=7.2$. We fitted both models with a PC having a 2.7 GHz 12-core Intel Xeon E5 processor and 64 GB of 1866~MHz DDR3 memory. Fitting our novel model in equation~\eqref{mod_model} required seventeen minutes and thirteen seconds, whereas fitting the model by \cite{Benko2018} in equation~\eqref{Benko_model} required
twelve minutes and twenty-eight seconds.

The description provided so far applies to both fits of models in equations \eqref{mod_model} and \eqref{Benko_model}.
We now provide, separately, computational details about the estimation of these two models. Then in Sections~\ref{acc_app} and \ref{tv_app} we compare and interpret the fits. 

To fit our novel model in equation~\eqref{mod_model}, we apply the methodology described in Section \ref{Sec:estimation}. When fitting our model in equation~\eqref{mod_model} we consider two sets of harmonic components. The first set is given by the harmonic components with frequencies $\{f_{\!_{k}}=k f_{\!_{0}},\, k=1,\dots,15\}$ provided by \cite{Benko2014}, weighted by our amplitudes $\{g_{\!_{\ell,k}}(t_i),\, \ell=1,2,\, k=1,\dots,15\}$. For the second set, the harmonic components are characterized by the four amplitudes  $\{g'_{\!_{\ell,j}}(t_i),\, \ell=1,2,\, j=11,\dots,14\}$ weighting the corresponding four frequencies $\{f'_{\!_{j}},\,j=11,\dots,14\}$. That is, we fit the following extended version 
\begin{eqnarray*}
\mu(t_i)=m(t_i)&+&\sum^{15}_{k=1}\{g_{\!_{1,k}}(t_i)\cos(w_{\!_{k}}t_i)+g_{\!_{2,k}}(t_i)\sin(w_{\!_{k}}t_i)\}\\
&+&\sum^{14}_{j=11}\{g'_{\!_{1,j}}(t_i)\cos(w'_{\!_{j}}t_i)+g'_{\!_{2,j}}(t_i)\sin(w'_{\!_{j}}t_i)\}
\end{eqnarray*}
of model in equation~\eqref{mod_model}, with $\omega_k=2\pi f_k$ and $\omega'_j=2\pi f'_j$.
The resulting fitted model involves a total of $39J$ parameters. Before fitting our model, we selected the smoothing parameters $\vect{\tau}$ and the number of $B$-splines $J$. These parameters were selected by the AIC criterion described in Section \ref{CV}. To simplify the selection of the smoothing parameters $\vect{\tau}$, we consider the case $\tau_{\!_{2}}=\dots=\tau_{\!_{11}}$, $\tau_{\!_{12}}=\dots=\tau_{\!_{21}}$, $\tau_{\!_{22}}=\dots=\tau_{\!_{31}}$, and $\tau_{\!_{32}}=\dots=\tau_{\!_{39}}$. We pick the smoothing parameter $\tau_{\!_{1}}$ over the grid $\{ 0,0.1,10 \}$, the parameters $\{\tau_{\!_{k}},\, k=2,\dots,39\}$ over the grid $\{ 0,0.1,10,100\}$, and the total number $J$ of $B$-splines (of degree $d=3$) over the grid $\{8,13,23,33\}$. We apply the AIC formula in equation~\eqref{AIC}. The lowest AIC value occurs for $J=33$ $B$-splines,  $\tau_{\!_{1}}=0$, $\tau_{\!_{k}}=0.1$, $k=2,\dots,21$, $\tau_{\!_{k}}=0$, $k=22,\dots,31$, and $\tau_{\!_{k}}=10$, $k=32,\dots,39$.

To fit the model in equation~\eqref{Benko_model}, we  implement the Levenberg–Marquardt algorithm using the {\tt R} function {\tt nls.lm}
\citep[available in the {\tt R} package {\tt minpack.lm} by][] {minpack2016}, with $K=15$ and  $\ell=\ell^A_{\!_{k}}=\ell^F_{\!_{k}}=1$, $k=1,\dots,K$, for a total of 93 parameters. 

\subsection{Comparing the accuracy of the fits}\label{acc_app}
The MSE corresponding to the fit of our model in equation~\eqref{mod_model} is 0.000001, whereas the MSE of the model in equation~\eqref{Benko_model} is 0.000008. That is, the MSE of the model in equation~\eqref{mod_model} is approximately 12.5\%  smaller than the MSE of the model in equation~\eqref{Benko_model}. Fitting the model in equation~\eqref{mod_model} involves 1,287 parameters, whereas the number of parameters estimated with the model in equation~\eqref{Benko_model} is 93. The larger number of parameters needed to fit the model in equation~\eqref{mod_model} is due to the semi-parametric form of trend and amplitudes, which does not impose any particular shape to the underlying functions we estimate.

\begin{figure}[htb!]
\centering
\includegraphics[width=18cm]{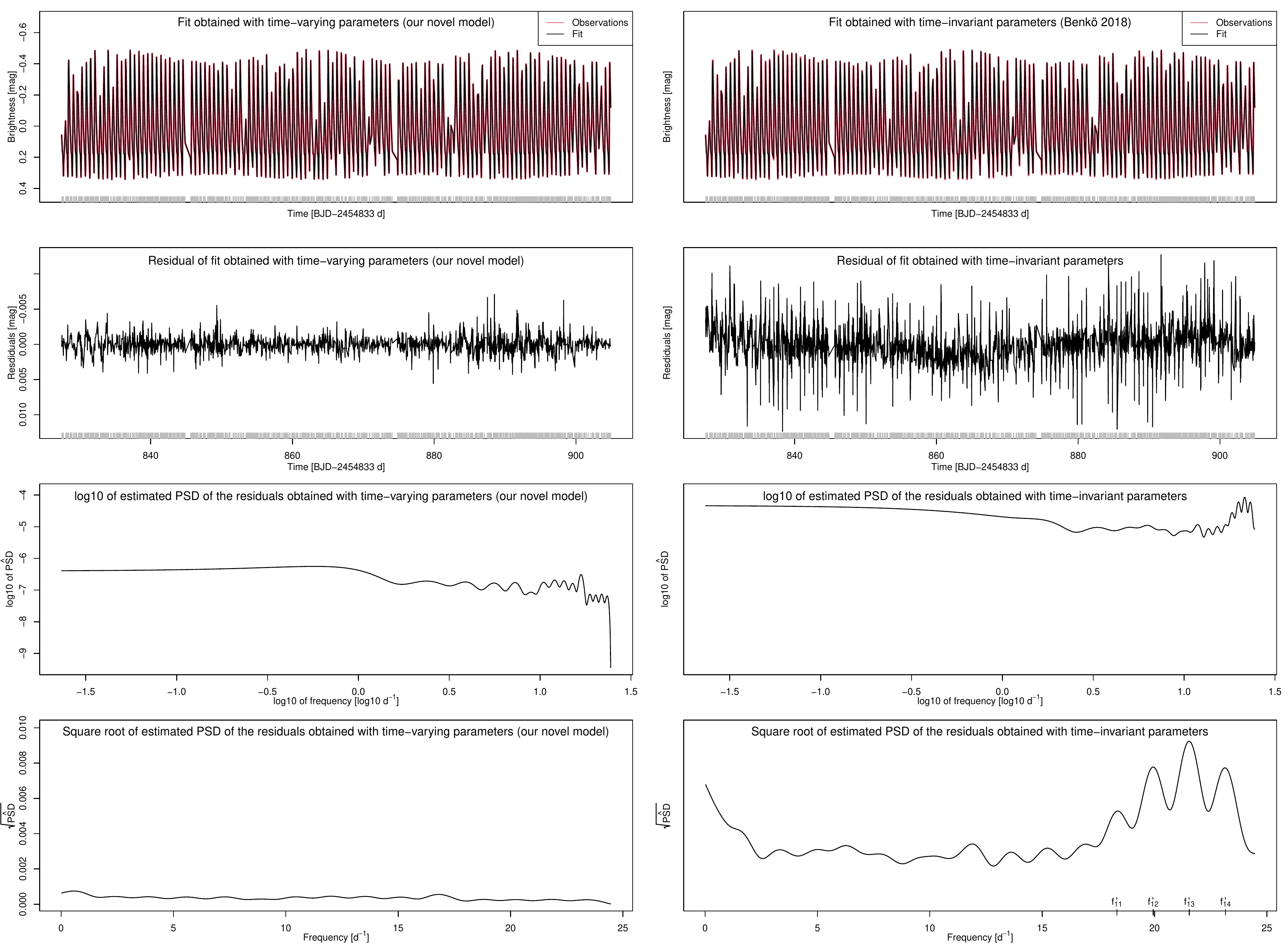}
\caption{Comparison of fitted models to the light curve V783 Cyg in Section \ref{acc_app}. The first column corresponds to the fit of the model in equation~\eqref{mod_model}, whereas the second column corresponds to the fit of the model in equation~\eqref{Benko_model}. From top to bottom: the first row shows the Brightness mag (red solid lines) together with the fits (black solid lines). The second row shows the residuals resulting from the fits. The last two rows show the  spectral density of the residual obtained with  equation~\eqref{PSD_hat} under different transformations (log-10 scale and square root).}\label{Figure:V783_fit}
\end{figure}

Figure~\ref{Figure:V783_fit} compares the fits of the model in equations \eqref{mod_model} and \eqref{Benko_model}. The first row shows the fitted curves, the second row shows the residuals, and the third and fourth rows show the estimated spectral density of the residuals. Albeit the fitted curves (first row) look very similar, the residuals are significantly different. Indeed, the residuals obtained with the model in equation~\eqref{mod_model} are compatible with the assumption of stationary and uncorrelated errors. By contrast, the residuals obtained with the model in equation~\eqref{Benko_model} exhibit time-dependent trend. Moreover, the estimated spectral densities in the last two rows of Figure~\ref{Figure:V783_fit} show that the model in equation~\eqref{mod_model} delivers residuals with a flat estimated spectral density, mimicking the behavior of the spectral density of white noise errors, whereas for the model in equation~\eqref{Benko_model} shows that some harmonic components should be added to the model (see the peaks between the frequencies 17 d$^{-1}$ and 24 d$^{-1}$ in the last row).

\subsection{Comparing the estimated time-varying parameters}\label{tv_app}
In Section \ref{Sec:modulation} we have showed that the model in equation~\eqref{Benko_model} is a special case of our novel model in equation~\eqref{mod_model}. To establish whether the fitted model in equation~\eqref{mod_model} matches (or differs from) the fitted model in equation~\eqref{Benko_model}, we now compare the estimates of $m(t)$ and $g_{\!_{\ell,k}}(t)$ obtained by fitting the model in equation~\eqref{mod_model} with the estimates of $u(t)$ and $h_{\!_{\ell,k}}(t)$ defined in equation~\eqref{Benko_comp} obtained by fitting the model in equation~\eqref{Benko_model}. 

\begin{figure}[htb!]
\centering
\includegraphics[width=18cm]{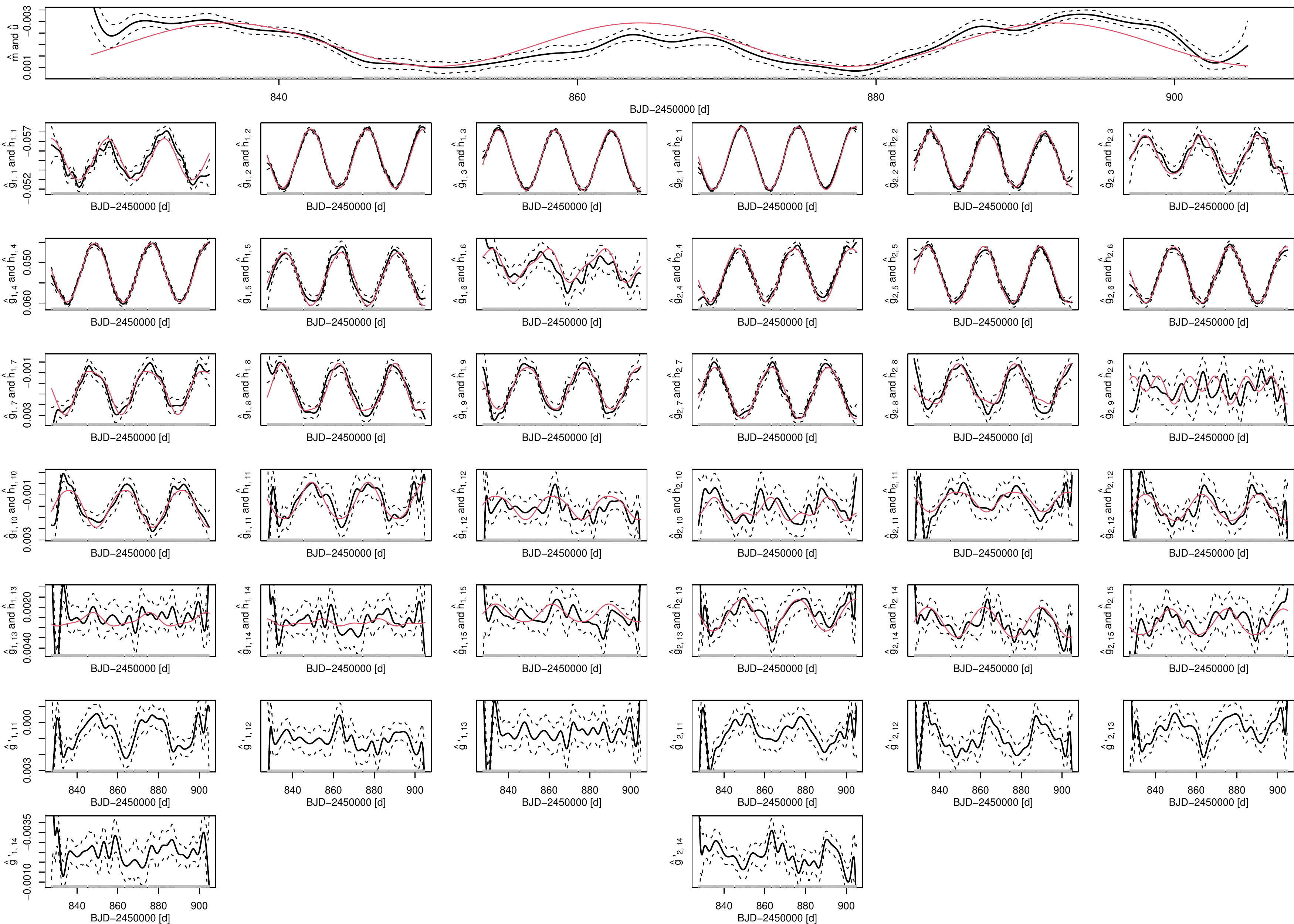}
\caption{Comparing the estimated time-varying trend and amplitudes fitted to the light curve V783 Cyg studied in Section \ref{tv_app}. Red solid lines: estimates of $u(t)$ and $\{h_{\!_{\ell,k}}(t),\, \ell=1,2,\, k=1,\dots,15\}$ defined in equation~\eqref{Benko_comp} obtained with the model in equation~\eqref{Benko_model}. Black solid lines: estimates of $m(t)$, $\{g_{\!_{\ell,k}}(t),\, \ell=1,2,\, k=1,\dots,15\}$,  and $\{g'_{\!_{\ell,j}}(t),\, \ell=1,2,\, j=11,\dots,14\}$ obtained with our novel model in equation~\eqref{mod_model}. Black dashed lines: 95\% confidence intervals for $m(t)$, $g_{\!_{\ell,k}}(t)$ and $g'_{\!_{\ell,j}}(t)$ obtained according to Appendix~\ref{Ape_B_CI_par}.}\label{Figure:V783_fit_g}
\end{figure}

Figure~\ref{Figure:V783_fit_g} shows the estimates $\widehat{m}(t)$, $\{\widehat{g}_{\!_{\ell,k}}(t),\, \ell=1,2,\, k=1,\dots,15\}$, and $\{\widehat{g}'_{\!_{\ell,j}}(t),\, \ell=1,2,\, j=11,\dots,14\}$ (black lines) , together with the estimates $\widehat{u}(t)$ and $\{\widehat{h}_{\!_{\ell,k}}(t),\, \ell=1,2,\, k=1,\dots,15\}$ (red lines). The estimated trend $\widehat{m}(t)$ is similar to the sinusoidal $\widehat{u}(t)$. Similarly, the first eight estimated harmonic components $\{\widehat{g}_{\!_{\ell,k}}(t),\,\ell=1,2,\, k=1,\dots,8\}$ and $\{\widehat{h}_{\!_{\ell,k}}(t),\,\ell=1,2,\, k=1,\dots,8\}$ are very close to each other. The next seven estimated harmonic components $\{\widehat{g}_{\!_{\ell,k}}(t),\,\ell=1,2,\, k=9,\dots,15\}$ and $\{\widehat{h}_{\!_{\ell,k}}(t),\,\ell=1,2,\, k=9,\dots,15\}$ 
are still similar but in some cases are slightly different.
Nevertheless, these small differences do not have a significant impact on the fitted curves, because the last harmonic components have less contribution to the fit than the first ones. For the four estimated harmonic components associated to the frequencies $\{f'_{\!_{j}},\, j=11,\dots,14\}$, which were fitted only for the model in equation~\eqref{mod_model} -- and were not fitted for the model in equation~\eqref{Benko_model} -- we observe that the four corresponding time-varying amplitudes $\{\widehat{g}'_{\!_{\ell,j}}(t),\, \ell=1,2,\, j=11,\dots,14\}$ are allowed to have either a sinusoidal or a non-sinusoidal form. This finding is in accordance with the form of Amplitude Modulation  and Frequency Modulation of Blazhko stars described by \cite{Benko2018}. Finally, in Figure~\ref{Figure:V783_fit_g}, we see that most of the confidence intervals of $\widehat{g}_{\!_{\ell,k}}(t)$ contain $\widehat{h}_{\!_{\ell,k}}(t)$. Therefore we conclude the following. Albeit the modulation frequency $f_{\!_{m}}$ is not a parameter of our model in equation~\eqref{mod_model}, we are able to describe, through the estimated time-varying trend $\widehat{m}(t)$ and amplitudes $\widehat{g}_{\!_{\ell,k}}(t)$, the Blazhko effect resulting from the amplitude and frequency modulation considered by the model in equation~\eqref{Benko_model}. 

\section{Summary}\label{Sec:summary}

In this article, we introduced a model for time series observations of variable stars that are modulated by smoothly time-varying mean magnitudes, amplitudes, and phases. Previous approaches assume that the underlying parameters are either time-invariant or piecewise-constant functions. From the modeling viewpoint, our approach is more flexible because it avoids assumptions about the functional form of the aforementioned time-dependent quantities. From the computational viewpoint, estimating our time-varying curves translates into the estimation of time-invariant parameters that can be performed by ordinary least-squares.

An important challenge when dealing with astronomical time series is that observations are unequally spaced in time. In some cases, observations are unevenly spaced due to missing values. Missing values are sometimes handled via imputation, that is, the gap generated by the missing value is ``filled in" by an estimated value. Our novel approach, which involves the classical periodogram, has the advantage of not relying on any imputation method. 

We study the performance of our approach under several simulation scenarios. Finally, we apply our method to V783~Cyg (KIC 5559631), a well-known RR Lyrae star presenting the Blazhko effect. In this case, the effect is characterized by a sinusoidal amplitude and frequency modulation. When comparing the time-varying fit obtained with our novel model with the time-invariant fit obtained with the  model proposed by \cite{Benko2018}, we found that both amplitude and frequency modulations are well captured and fitted by our novel model, and also that our time-varying method outperforms the time-invariant fit. Indeed the estimation error obtained with our fit is significantly smaller than the error obtained with the time-invariant fit. In addition, the residuals obtained with our novel method are compatible with the assumption of stationary and uncorrelated errors, whereas the residuals obtained with the time-invariant model by \cite{Benko2018} exhibit a time-dependent trend and some significant spectral peaks. 

In the future, we plan to extend our methodology in four important directions. First, we plan to apply our novel method to the study of a larger sample of Blazhko RR Lyrae stars. Second, our approach can be extended to the analysis of other classes of variable stars presenting long-term changes in their light curve shapes. Third, our fitting method does not require the period(s), amplitude(s), and phase(s) of the Blazhko effect to be determined, as we obtain instead the empirical functions $m(\cdot)$ and $g_{i,k}(\cdot)$. We are currently investigating what kind of (or how much more) information can be obtained from these empirical functions, as compared to conventional approaches. Finally, we aim to study Blazhko light curves characterized by more than one Blazhko frequency~--  V783 Cyg, which was addressed in some detail in this paper, is a special case, because this star does not show any additional Blazhko frequencies  \citep[][]{Benko2014}.

\begin{acknowledgments}
The authors would like to thank two anonymous reviewers for helpful comments which led to significant improvements of the paper.
D.S. was funded by the National Agency for Research and Development (ANID), Doctorado Nacional grant 2017-21171100.
Support for G.M. and M.C. has been provided by ANID's Millennium Science Initiative through grant ICN12\textunderscore 120009, awarded to the Millennium Institute of Astrophysics (MAS). M.C. acknowledges additional support by Proyecto Basal AFB-170002 and FONDECYT grant \#1171273.
We thank all the participants of the Astronomical Data Science Workshop organized by Texas A\&M University on February 17-18, 2020, as well as the participants of the IISA 2021 conference organized by the University of Illinois Chicago on May 20-23, 2021. We thank the Astrostatistics group at the Center for Astrophysics of Harvard University for constructive criticism about this manuscript. 
Special thanks go to J\'ozsef {Benk{\H{o}}} for sharing the parameters we used to simulate the Blazhko star of  Section~\ref{sim_blazhko}, and to Gergely Hajdu for useful discussions. 
\end{acknowledgments}

\software{ The Language R for Statistical Computing \citep{R2021}, the R package {\tt NISTunits} \citep{NISTunits2016}, and the R package {\tt minpack.lm} \citep{minpack2016}. }
          
\newpage

\appendix

\section{Modulation} \label{Ape_modulation}
The Blazhko effect is a periodic amplitude and phase variation in the light curves of RR Lyrae variable stars. In astronomy, the Blazhko effect is usually interpreted as a modulation phenomenon. Modulation is the process of transmitting a low-frequency signal into a high-frequency wave, called the \textit{carrier wave}, by changing its amplitude, frequency, and/or phase angle through the \textit{modulating signal}. The function of the carrier wave is to carry the message or modulating signal from the transmitter to the receiver. The superposition of the signal and the carrier wave results in the so-called \textit{modulated signal}.

In this Appendix we review two types of modulation, as given in \cite{Benko2011}: amplitude modulation and frequency modulation. This will be helpful for a comparison between our model (eq.~\ref{mod_model}) and the models proposed by \cite{Benko2011} and \cite{Benko2018}, in the case of 
RR Lyrae stars presenting the Blazhko effect. 

\subsection{Amplitude modulation}

Amplitude modulation (AM) changes the amplitude of the carrier signal. Let the carrier wave $c(t)$ be a sinusoidal signal of the form
\begin{equation*}
c(t)=U_{\!_{c}} \sin(2 \pi f_{\!_{c}} t+\phi_{\!_{c}}),
\end{equation*}
where the constant parameters $U_{\!_{c}}$, $f_{\!_{c}}$, and $\phi_{\!_{c}}$ are the amplitude, frequency, and phase of the carrier wave, respectively.

Let $U_{\!_{m}}(t)$ represent a waveform that is the message to be transmitted, or \textit{modulating} signal. The transmitter uses the information signal $U_{\!_{m}}(t)$ to vary the amplitude of the carrier $U_{\!_{c}}$ to produce the \textit{amplitude modulated} signal $U_{\!_{\text{AM}}}$:
\begin{equation}\label{AM1}
U_{\!_{\text{AM}}}(t)=[U_{\!_{c}}+U_{\!_{m}}(t)]\sin(2 \pi f_{\!_{c}} t+\phi_{\!_{c}})
=[U_{\!_{c}}+U_{\!_{m}}(t)]\tfrac{c(t)}{U_{\!_{c}}}= \left[1+ \tfrac{U_{\!_{m}}}{U_{\!_{c}}} \right] c(t).
\end{equation}

\noindent In the simplest case, when the modulating signal is sinusoidal, that is,
\begin{equation}\label{AM7}
U_{\!_{m}}(t)=U^{A}_{\!_{m}} \sin (2 \pi f_{\!_{m}} t + \phi^{A}_{\!_{m}}),
\end{equation}
the amplitude-modulated signal in equation \eqref{AM1} is
\begin{equation}\label{AM5}
U_{\!_{\text{AM}}}(t)=\left[U_{\!_{c}}+U^{A}_{\!_{m}} \sin (2 \pi f_{\!_{m}} + \phi^{A}_{\!_{m}})\right]\sin(2 \pi f_{\!_{c}} t+\phi_{\!_{c}}).
\end{equation}

Clearly, a more complex example of amplitude modulation arises when $K\geq 1$, where $K$ denotes the number of harmonic components.
Suppose the carrier wave $c(t)$ is a linear combination of sine harmonics: 
\begin{equation*}
c(t)=a_{\!_{0}}+\sum^K_{k=1} a_{\!_{k}} \sin(2 \pi k f_{\!_{0}} t+\phi_{\!_{k}}),
\end{equation*}
and the modulating signal is sinusoidal and given again by equation \eqref{AM7}.
  Following the same idea as in equation \eqref{AM1}, the amplitude-modulated signal in equation \eqref{AM5} is
\begin{align}\label{AM8}
\begin{split}
U_{\!_{\text{AM}}}(t)&=
\left[1+\frac{U_{\!_{m}}(t)}{U_{\!_{c}}} \right ]\,c(t)\\
&=\left [ 1+\frac{U^A_{\!_{m}}\sin(2 \pi f_{\!_{m}} t+\phi^A_{\!_{m}})}{U_{\!_{c}}} \right ]\left [a_{\!_{0}}+\sum^K_{k=1} a_{\!_{k}} \sin(2 \pi k f_{\!_{0}} t+\phi_{\!_{k}})\right ].
\end{split}
\end{align}
If we call $h=U^A_{\!_{m}}/U_{\!_{c}}$, and use the basic trigonometrical identities $\sin(a)\sin(b)=\frac{1}{2}[\cos(a-b)-\cos(a+b)]$ and $\sin(a)=\cos(a-\frac{\pi}{2})$, equation \eqref{AM8} can be written as
\begin{align}\label{AM9}
\begin{split}
U_{\!_{\text{AM}}}(t)&=a_{\!_{0}}+\sum^K_{k=1} a_{\!_{k}} \sin(2 \pi k f_{\!_{0}} t+\phi_{\!_{k}})+a_{\!_{0}} h \sin(2 \pi f_{\!_{m}} t+\phi^A_{\!_{m}})\\ 
&+\sum^K_{k=1} \frac{a_{\!_{k}} h}{2} \sin(2 \pi (k f_{\!_{0}}-f_{\!_{m}}) t+(\phi_{\!_{k}}-\phi_{\!_{m}})+\pi/2)\\ 
&-\sum^K_{k=1} \frac{a_{\!_{k}} h}{2} \sin(2 \pi (k f_{\!_{0}}+f_{\!_{m}}) t+(\phi_{\!_{k}}+\phi_{\!_{m}})+\pi/2).
\end{split}
\end{align}

This example shows that when the time-varying amplitude $U_{\!_{m}}(t)$ in equation \eqref{AM7} takes a sinusoidal form, the amplitude modulated model with \textit{time-varying} amplitude in equation \eqref{AM8} can be written as a model with \textit{time-invariant} parameters as in equation \eqref{AM9}. This implies that, when frequencies and phases are known, the parameters $\{a_k,\,0\leq k\leq K\}$ in equation \eqref{AM9} can be estimated by ordinary least squares.

\subsection{Amplitude and frequency modulation}

Frequency modulation (FM) changes the frequency of the carrier signal. We assume the sinusoidal carrier wave to be
\begin{equation*}
c(t)=U_{\!_{c}} \sin(\Theta(t)),
\end{equation*}
where $\Theta(t)=2\pi f_{\!_{c}} t+\phi_{\!_{c}}$ is the angular part of the function. Suppose that the modulating signal is $U_{\!_{m}}(t)$. Then the modulated angular part is given by
\begin{equation*}
\Theta(t)= 2\pi f_{\!_{c}} t + 2\pi k_{\!_{\text{FM}}} \int^t_0 U^{F}_{\!_{m}}(\tau) \text{d} \tau + \phi_{\!_{c}},
\end{equation*}
where $k_{\!_{\text{FM}}}$ is the frequency deviation, and the \textit{frequency modulated} signal is expressed as
\begin{equation}\label{FM1}
U_{\!_{\text{FM}}}(t)=U_{\!_{c}} \sin \left ( 2\pi f_{\!_{c}}t + 2\pi k_{\!_{\text{FM}}} \int^t_0 U^{F}_{\!_{m}}(\tau) \text{d} \tau +\phi_{\!_{c}} \right).
\end{equation}
In the simplest case, when the modulating signal is represented by a sinusoidal wave with amplitude $U^{F}_{\!_{m}}$ and frequency $f_{\!_{m}}$, the integral of such a signal is
$$\int^t_{0} U_{\!_{m}}(\tau) \text{d} \tau = \frac{U^{F}_{\!_{m}}}{2 \pi f_{\!_{m}}} \sin(2 \pi f_{\!_{m}} t + \phi_{\!_{m}}),$$
and the frequency-modulated signal in equation \eqref{FM1} is
\begin{equation}\label{FM2}
U_{\!_{\text{FM}}}(t)=U_{\!_{c}} \sin \left ( 2\pi f_{\!_{c}}t + \frac{k_{\!_{\text{FM}}} U^{F}_{\!_{m}}}{f_{\!_{m}}} \sin(2 \pi f_{\!_{m}} t + \phi_{\!_{m}}) +\phi_{\!_{c}} \right ).
\end{equation}
In practice, modulated signals can be a mixture of amplitude and frequency modulations, which can be used to described Blazhko RR Lyrae stars \citep{Benko2011}. We review the simplest case when both AM and FM are sinusoidal. Combining the amplitude modulated signal in equation \eqref{AM5} and the frequency modulated signal in equation \eqref{FM2}, the \textit{amplitude and frequency modulated} signal is thus
\begin{equation*}
U_{\!_{\text{Comb}}}(t)=[U_{\!_{c}} + U^{A}_{\!_{m}} \sin(2 \pi f_{\!_{m}} t + \phi_{\!_{m}})] \sin \left ( 2 \pi f_{\!_{c}} t + \frac{k_{\!_{\text{FM}}} U^{F}_{\!_{m}}}{f_{\!_{m}}} \sin(2 \pi f_{\!_{m}} t + \phi_{\!_{m}}) + \phi_{\!_{c}} \right ).
\end{equation*}

\subsection{Blazhko modulation}

Amplitude and frequency modulations have been observed in Blazhko RR Lyrae stars \citep[e.g.,][]{Chadid2010,Benko2010,Poretti2010,Sodor2012}. Assuming that the observed data sets are precise and long enough, \cite{Benko2011} proposed an amplitude and frequency modulation model for Blazhko RR Lyrae stars given by
\begin{equation}\label{Benko_ec}
m^*_{\!_{\text{Comb}}}(t)=\frac{m^*_{\!_{\text{AM}}}(t)}{c^*(t)}m^*_{\!_{\text{FM}}}(t),
\end{equation}
where $c^*(t)$ is the carrier wave, and the functions $m^*_{\!_{\text{AM}}}(t)$ and $m^*_{\!_{\text{FM}}}(t)$ are the non-sinusoidal amplitude and frequency modulations, given respectively by
\begin{align}
m^*_{\!_{\text{AM}}}(t)&=\left [ a^A_{\!_{0}} + \sum^{q}_{p=1} a^A_{\!_{p}} \sin(2\pi p f_{\!_{m}} t + \varphi^A_{\!_{p}}) \right ] c^*(t), \label{AM_ec}\\
m^*_{\!_{\text{FM}}}(t)&=a_{\!_{0}} + \sum^K_{k=1} a_{\!_{k}} \sin \left [ 2 \pi k f_{\!_{0}} t+ k a^F_{\!_{0}} + k \sum^{q}_{p=1} a^F_{\!_{p}} \sin(2 \pi p f_{\!_{m}} t + \varphi^F_{p}) + \varphi_{k} \right ] \label{FM_ec}.
\end{align}
Here, the modulating signal used in the amplitude and frequency modulation is an arbitrary periodic signal represented by a Fourier sum with a
constant frequency $f_{\!_{m}}$. Superscripts A and F denote the amplitude modulation and frequency modulation parameters, respectively, and $f_{\!_{0}}$ and $f_{\!_{m}}$ are the main pulsation and modulation frequencies, respectively.

Substituting equations \eqref{AM_ec} and \eqref{FM_ec} in equation \eqref{Benko_ec}, we finally have
\begin{equation*}
\small{
m^*_{\!_{\text{Comb}}}(t)=
\left [ a^A_{\!_{0}} + \sum^{q}_{p=1} a^A_{\!_{p}} \sin(2 \pi p f_{\!_{m}} t + \varphi^A_{\!_{p}} ) \right ] \times \left \{ a_{\!_{0}} + \sum^K_{k=1} a_{\!_{k}} \sin \left [ 2 \pi k f_{\!_{0}} t + k a^F_{\!_{0}} + k \sum^q_{p=1} a^F_{\!_{p}} \sin(2 \pi p f_{\!_{m}} t + \varphi^F_{\!_{p}} ) + \varphi_{\!_{k}} \right ] \right \}.
}
\end{equation*}

\section{$B$-splines} \label{Ape_B_splines}
In this Appendix we define $B$-splines and give some details about the estimation method that we used in this manuscript. For more details we refer the reader to the book by \cite{boor1978}.

A $B$-spline curve $f(t)$ of degree $d$ is defined as
\begin{equation}\label{def_B_spline_curve}
f(t)=\sum^{J}_{j=1} P_{\!_{j}} B_{\!_{j,d}}(t),
\end{equation}
where $P_{\!_{j}}$ are the control points and $B_{\!_{j,d}}(t)$ are the $B$-spline basis functions. Let $t_{\!_{\rm min}}$ and $t_{\!_{\rm max}}$ be, respectively, the lower and upper bounds of the domain of interest. In order to build the $B$-spline basis of degree $d$, we firstly divide the domain into $n$ intervals, with $n$ being a positive  integer, obtaining the $n+1$ knots $\xi_{\!_{\,d}},\xi_{\!_{\,d+1}},\dots,\xi_{\!_{\,d+n}}$. Each knot satisfies $\xi_{\!_{\,j}}<\xi_{\!_{\,j+1}}$, for all $j$. Secondly, we define $2d$ additional knots $\xi_{\!_{\,0}},\xi_{\!_{\,1}},\dots,\xi_{\!_{\,d-1}},\xi_{\!_{\,n+d+1}},\dots,\xi_{\!_{\,n+2d-1}},\xi_{\!_{\,n+2d}}$. Then, the $j$th $B$-spline basis, $B_{\!_{j,d}}(t)$, can be defined recursively as 
\begin{eqnarray}\label{def_B_spline}
B_{\!_{j,d}}(t)=\frac{t-\xi_{\!_{\,j-1}}}{\xi_{\!_{\,j+d-1}}-\xi_{\!_{\,j-1}}} B_{\!_{j,d-1}}(t) + \frac{\xi_{\!_{\,j+d}}-t}{\xi_{\!_{\,j+d}}-\xi_{\!_{\,j}}} B_{\!_{j+1,d-1}}(t), \quad j=1,\dots,J,
\end{eqnarray}
with
\begin{equation}\label{B_spline_j_1}
B_{\!_{j,0}}(t) =
  \begin{cases}
    1 & t \in [\xi_{\!_{\,j-1}},\xi_{\!_{\,j}}), \\
    0 & \text{otherwise}
  \end{cases}
\end{equation}
being used to initialize the recursion. Thus, to build the $B$-spline curve given by equation \eqref{def_B_spline_curve}, we need $n+2d+1$ knots, and the total number of $B$-splines basis functions is $J=n+d$. 

To illustrate how to construct a $B$-spline basis, consider the case of degree $d=2$ and assume that the domain $[t_{\!_{min}},t_{\!_{max}}]$ has been divided into $n=3$ intervals, obtaining the knots $\xi_{\!_{\,2}},\dots,\xi_{\!_{\,5}}$. In this instance, the $2d=4$ additional knots are defined as $\xi_{\!_{\,0}},\xi_{\!_{\,1}},\xi_{\!_{\,6}},\xi_{\!_{\,7}}$. Using equation \eqref{def_B_spline}, we obtain
\begin{align*}
B_{\!_{1,2}}(t) & =\frac{t-\xi_{\!_{\,0}}}{\xi_{\!_{\,2}}-\xi_{\!_{\,0}}} B_{\!_{1,1}}(t) + \frac{\xi_{\!_{\,3}}-t}{\xi_{\!_{\,3}}-\xi_{\!_{\,1}}} B_{\!_{2,1}}(t), \\
B_{\!_{2,2}}(t) & =\frac{t-\xi_{\!_{\,1}}}{\xi_{\!_{\,3}}-\xi_{\!_{\,1}}} B_{\!_{2,1}}(t) + \frac{\xi_{\!_{\,4}}-t}{\xi_{\!_{\,4}}-\xi_{\!_{\,2}}} B_{\!_{3,1}}(t), \\
B_{\!_{3,2}}(t) & =\frac{t-\xi_{\!_{\,2}}}{\xi_{\!_{\,4}}-\xi_{\!_{\,2}}} B_{\!_{3,1}}(t) + \frac{\xi_{\!_{\,5}}-t}{\xi_{\!_{\,5}}-\xi_{\!_{\,3}}} B_{\!_{4,1}}(t), \\
B_{\!_{4,2}}(t) & =\frac{t-\xi_{\!_{\,3}}}{\xi_{\!_{\,5}}-\xi_{\!_{\,3}}} B_{\!_{4,1}}(t) + \frac{\xi_{\!_{\,6}}-t}{\xi_{\!_{\,6}}-\xi_{\!_{\,4}}} B_{\!_{5,1}}(t), \\
B_{\!_{5,2}}(t) & =\frac{t-\xi_{\!_{\,4}}}{\xi_{\!_{\,6}}-\xi_{\!_{\,4}}} B_{\!_{5,1}}(t) + \frac{\xi_{\!_{\,7}}-t}{\xi_{\!_{\,7}}-\xi_{\!_{\,5}}} B_{\!_{6,1}}(t),
\end{align*}
where 
\vspace{-.3cm}
\begin{align*}
B_{\!_{1,1}}(t) & =\frac{t-\xi_{\!_{\,0}}}{\xi_{\!_{\,1}}-\xi_{\!_{\,0}}} B_{\!_{1,0}}(t) + \frac{\xi_{\!_{\,2}}-t}{\xi_{\!_{\,2}}-\xi_{\!_{\,1}}} B_{\!_{2,0}}(t), \\
B_{\!_{2,1}}(t) & =\frac{t-\xi_{\!_{\,1}}}{\xi_{\!_{\,2}}-\xi_{\!_{\,1}}} B_{\!_{2,0}}(t) + \frac{\xi_{\!_{\,3}}-t}{\xi_{\!_{\,3}}-\xi_{\!_{\,2}}} B_{\!_{3,0}}(t), \\
B_{\!_{3,1}}(t) & =\frac{t-\xi_{\!_{\,2}}}{\xi_{\!_{\,3}}-\xi_{\!_{\,2}}} B_{\!_{3,0}}(t) + \frac{\xi_{\!_{\,4}}-t}{\xi_{\!_{\,4}}-\xi_{\!_{\,3}}} B_{\!_{4,0}}(t), \\
B_{\!_{4,1}}(t) & =\frac{t-\xi_{\!_{\,3}}}{\xi_{\!_{\,4}}-\xi_{\!_{\,3}}} B_{\!_{4,0}}(t) + \frac{\xi_{\!_{\,5}}-t}{\xi_{\!_{\,5}}-\xi_{\!_{\,4}}} B_{\!_{5,0}}(t), \\
B_{\!_{5,1}}(t) & =\frac{t-\xi_{\!_{\,4}}}{\xi_{\!_{\,5}}-\xi_{\!_{\,4}}} B_{\!_{5,0}}(t) + \frac{\xi_{\!_{\,6}}-t}{\xi_{\!_{\,6}}-\xi_{\!_{\,5}}} B_{\!_{6,0}}(t), \\
B_{\!_{6,1}}(t) & =\frac{t-\xi_{\!_{\,5}}}{\xi_{\!_{\,6}}-\xi_{\!_{\,5}}} B_{\!_{6,0}}(t) + \frac{\xi_{\!_{\,7}}-t}{\xi_{\!_{\,7}}-\xi_{\!_{\,6}}} B_{\!_{7,0}}(t),
\end{align*}
and the coefficients $\{ B_{\!_{j,0}}(t)$, $j=1,\dots,7\}$ are defined in equation \eqref{B_spline_j_1}.

Suppose we have $N$ observations $\{t_1,\dots,t_N\}$, that might be either equally or unequally spaced, with $t_i \in [t_{\!_{min}},t_{\!_{max}}]$ for all $i=1,\dots,N$. The $B$-splines basis matrix evaluated at time $\{t_1,\dots,t_N\}$, denoted by $\matr{B}$, is the $N\times J$ matrix with entries $\{B_{\!_{j,d}}(t_i),\,i=1,\dots,N,\,j=1,\dots,J\}$, in a way that each row contains a B-spline basis. The $j$th $B$-spline basis function satisfies the following properties:
\[
\begin{cases}
B_{\!_{j,d}}(t)>0 & \xi_{\!_{\,j-1}} < t < \xi_{\!_{\,j+d}},\\
B_{\!_{j,d}}(t)=0& \xi_{\!_{\,0}} \leq t \leq \xi_{\!_{\,j-1}}\,\,\mbox{and}\,\,\xi_{\!_{\,j+d}} \leq t \leq \xi_{\!_{\,n+2d}},\\
\sum^J_{j=1}B_{\!_{j,d}}(t)=1 & \xi_{\!_{\,d}} < t < \xi_{\!_{\,n+d}},\\
\tfrac{\partial^k B_{\!_{j,d}}(t)}{\partial t^k}|_{t= \xi_{\!_{\ell}}} & 0\leq k\leq d-1 \,\,\mbox{are continuous}.\\
\end{cases}
\]
For ease of notation, we use, throughout our manuscript, $B_{\!_{j}}(t)$ instead of $B_{\!_{j,d}}(t)$. Let us now consider the example of estimating the mean function $\mu(t)$ of model 
$Y_{\!_{i}}=\mu(t_i)+\varepsilon_{\!_{i}}$
using $B$-splines. Let $\vect{Y}=(Y_{\!_{1}},\dots,Y_{\!_{6}})'$ be the available $N=6$ responses observed, respectively, at time $\{t_1,\dots,t_6\}$, with $t_{\!_{min}}=t_1$ and $t_{\!_{max}}=t_6$. Then assume that $\mu(t)=\sum^{J}_{j=1} P_{\!_{j}} B_{\!_{j}}(t)$, for all $t \in [t_1,t_6]$. We use here $B$-splines basis functions of degree $d=2$; in order to construct them, we divide the domain $[t_1,t_6]$ into $n=3$ intervals. Hence, the total number of knots $\xi_{\!_{\,0}},\dots,\xi_{\!_{\,7}}$ is $n+2d+1=8$, and the total number of $B$-splines basis functions is $J=n+d=5$. The $6\times 5$ design matrix $\matr{B}$ has entries $B_{ij}=B_{\!{j}}(t_i)$, with $i=1,\dots,6$ and $j=1,\dots,5$, which permits estimating the coefficients $\{P_{\!_{j}},\,j=1,\dots,5\}$ by ordinary least squares. Indeed, if $\vect{Y}=(Y_{\!_{1}},\dots,Y_{\!_{6}})\tp$ 
denotes the  response-vector and $\vect{\theta}=(P_{\!_{1}},\dots,P_{\!_{5}})\tp$ the parameter-vector, we can rewrite the model as $\vect{Y} = \matr{B}\vect{\theta} +\vect{z}$, where $\vect{z}=(z_{\!_1},\dots,z_{\!_6})\tp$ is the error vector. The estimated parameters are defined as $\widehat{\vect{\theta}}=(\widehat{P}_{\!_{1}},\dots,\widehat{P}_{\!_5})\tp=(\matr{B}\tp \matr{B})^{-1}\matr{B}\tp\vect{Y}$, and the estimated mean as $\widehat{\mu}(t)=\sum^{5}_{j=1} \widehat{P}_{\!_{j}} B_{\!_{j}}(t)$, for all $t \in [t_1,t_6]$. 

\section{Confidence Intervals}\label{app_CI}
\subsection{Non-parametric quantiles}\label{Ape_B_CI_nonpar}
We use the quantiles 0.025 and 0.975 to construct the confidence intervals in our simulations of Section~\ref{sim_tv_model}. For $t$ fixed, confidence intervals for $\mu(t)$, $m(t)$, and ${g}_{\!_{\ell,k}}(t)$, $\ell=1,2$, $k=1,\dots,K$, are calculated according to the following 3 steps:

\begin{enumerate}
\item We estimate the coefficients of interest $\vect{\alpha}$, $\vect{\beta}_{\!_{k}}$ and $\vect{\gamma}_{\!_{k}}$, $k=1,\dots,K$, following the procedure described in Section~\ref{sec:PLS}, and we define the $J \times M$ matrices $\matr{A}_{\!_{M}}=\left [ \widehat{\vect{\alpha}}^{(1)},\dots,\widehat{\vect{\alpha}}^{(M)} \right ]$, $\matr{B}_{\!_{kM}}=\left [\widehat{\vect{\beta}}^{(1)}_{\!_{k}},\dots,\widehat{\vect{\beta}}^{(M)}_{\!_{k}} \right ]$, and 
$\matr{G}_{\!_{kM}}=\left  [\widehat{\vect{\gamma}}^{(1)}_{\!_{k}},\dots,\widehat{\vect{\gamma}}^{(M)}_{\!_{k}}\right]$, where $\widehat{\vect{\alpha}}^{(j)}$, $\widehat{\vect{\beta}}^{(j)}_{\!_{k}}$, and  $\widehat{\vect{\gamma}}^{(j)}_{\!_{k}}$, correspond to the estimators of $\vect{\alpha}$, $\vect{\beta}_{\!_{k}}$, and $\vect{\gamma}_{\!_{k}}$, given by equation \eqref{theta_hat} in the $j$th Monte Carlo simulation, $j=1,\dots,M$.

\item We define the $M\times 1$ vectors $\widehat{\vect{m}}^{(M)}(t)$,  $\widehat{\vect{g}}^{(M)}_{\!_{\ell,k}}(t)$, $\ell=1,2$, $k=1,\dots,K$, and $\widehat{\vect{\mu}}^{(M)}(t)$ as 
\begin{align*}
\begin{split}
\widehat{\vect{m}}^{(M)}(t)&=\vect{B}(t)\tp \matr{A}_{\!_{M}}=\left [\widehat{m}^{(1)}(t),\dots,\widehat{m}^{(M)}(t) \right ],\\ 
\widehat{\vect{g}}^{(M)}_{\!_{1,k}}(t)&=\vect{B}(t)\tp \matr{B}_{\!_{kM}}=\left [\widehat{g}^{(1)}_{\!_{1,k}}(t),\dots,\widehat{g}^{(M)}_{\!_{1,k}}(t) \right ], \qquad k=1,\dots,K\\
\widehat{\vect{g}}^{(M)}_{\!_{2,k}}(t)&=\vect{B}(t)\tp \matr{G}_{\!_{kM}}=\left [ \widehat{g}^{(1)}_{\!_{2,k}}(t),\dots,\widehat{g}^{(M)}_{\!_{2,k}}(t) \right ],  \qquad k=1,\dots,K \\
\widehat{\vect{\mu}}^{(M)}(t) &=\widehat{\vect{m}}(t) + \sum^K_{k=1} \left \{ \widehat{\vect{g}}_{\!_{1,k}}(t) \cos(w_{\!_{k}}t) + \widehat{\vect{g}}_{\!_{2,k}}(t)\sin(w_{\!_{k}}t)\right \}=\left [ \widehat{\mu}^{(1)}(t),\dots,\widehat{\mu}^{(M)}(t) \right ],
\end{split}
\end{align*}
where $\widehat{\mu}^{(j)}(t)$, $\widehat{m}^{(j)}(t)$, and   $\widehat{g}^{(j)}_{\!_{\ell,k}}(t/t_N)$, $\ell=1,2$, $k=1,\dots,K$ correspond to the estimators of $\mu(t)$, $m(t)$, $g_{\!_{\ell,k}}(t/t_N)$, given by equations \eqref{pred_POLS} and  \eqref{m_g_hat} in the $j$th Monte Carlo simulation, and $\vect{B}(t)$ corresponds to the vector formed by the $B$-splines evaluated at time $t$.

\item We calculate the empirical quantiles of order 0.025 and 0.975 of the $M\times 1$ vectors $\widehat{\vect{m}}^{(M)}(t)$, $\widehat{\vect{g}}^{(M)}_{\!_{\ell,k}}(t)$, $\ell=1,2$, $k=1,\dots,K$,  and $\widehat{\vect{\mu}}^{(M)}(t)$.\\
\end{enumerate}

\subsection{Parametric quantiles}\label{Ape_B_CI_par}
We use the parametric quantiles to construct the confidence intervals for our simulation in Section~\ref{sim_blazhko} and our application in Section~\ref{Sec:appl}.
Assuming that the error terms $\{z_{\!_{i}},\, i=1,\dots,N \}$ follow a Gaussian distribution with zero mean and variance $\sigma^2_{\!{z}}$, the $(1-\alpha)\times 100 \%$ prediction interval for $\mean{\widehat{Y}_{\!_{i}}}=\vect{\mathcal{B}}\tp(t_i) \mean {\widehat{\vect{\theta}}}$, with $i=1,\dots,N$, is 
$$\vect{\mathcal{B}}\tp(t_i) \widehat{\vect{\theta}} \pm z(1-\alpha /2) \sqrt{ \vect{\mathcal{B}}\tp(t_i)  \var{\widehat{\vect{\theta}}} \vect{\mathcal{B}}(t_i)},$$ where $z(1-\alpha /2)$ denotes the $(1-\alpha /2)$ quantile of the standard Gaussian distribution, and 
$$\var{\widehat{\vect{\theta}}} =\sigma^2_{\!{z}} ( \matr{\mathcal{B}} \tp  \matr{\mathcal{B}} + \matr{P})^{-1}\matr{\mathcal{B}} \tp \matr{\mathcal{B}} ( \matr{\mathcal{B}} \tp \matr{\mathcal{B}} + \matr{P})^{-1}.$$
The $(1-\alpha)\times 100 \%$ confidence interval for the trend  $\mean{\widehat{m}(t_i)}=\vect{\mathscr{B}}(t_i)\tp \matr{Q}_m\mean {\widehat{\vect{\theta}}}$ is
\begin{equation*}
\label{CI_m}
\vect{\mathscr{B}}(t_i)\tp \matr{Q}_m \widehat{\vect{\theta}} \pm z(1-\alpha /2) \sqrt{ \vect{\mathscr{B}}(t_i)\tp \matr{Q}_m \var{\widehat{\vect{\theta}}} \matr{Q}_m \tp \vect{\mathscr{B}}(t_i)},
\end{equation*}
and the $(1-\alpha)\times 100 \% $ confidence intervals for the amplitudes $\mean{\widehat{g}_{\!_{\ell,k}}(t_i)}=\vect{\mathscr{B}}(t_i)\tp \matr{Q}_g(\ell,k)\mean{\widehat{\vect{\theta}}}$, $\ell=1,2$, $k=1,\dots,K$, are
\begin{equation*}\label{CI_g}
\vect{\mathscr{B}}(t_i)\tp \matr{Q}_g(\ell,k)\widehat{\vect{\theta}}  \pm z(1-\alpha /2) \sqrt{ \vect{\mathscr{B}}(t_i) \tp \matr{Q}_g(\ell,k) \var{\widehat{\vect{\theta}}} \matr{Q}_g(\ell,k) \tp \vect{\mathscr{B}}(t_i)},   
\end{equation*}
where $\vect{\mathscr{B}}(t_i)$ is the $i$th row of the matrix $\matr{\mathscr{B}}$, and $\matr{\mathscr{B}}= [\matr{B}|\matr{B}|\dots|\matr{B}]$ is a matrix of dimension $N \times c$. The $c \times c$ matrices $\matr{Q}_m$, $\{\matr{Q}_g(\ell,k),\, \ell=1,2,\, k=1,\dots,K\}$ satisfy $\matr{Q}_m {\vect{\theta}} =(\vect{\alpha}\tp,\vect{0}_{\!_{J}}\tp,\dots,\vect{0}_{\!_{J}}\tp)\tp$ and 

\begin{align*}
\matr{Q}_g(1,1) {\vect{\theta}} 
& =(\vect{0}_{\!_{J}}\tp,\vect{\beta}_{\!_{1}}\tp,\vect{0}_{\!_{J}}\tp,\dots,\vect{0}_{\!_{J}}\tp)\tp, 
&\matr{Q}_g(2,1) {\vect{\theta}}
& =(\vect{0}_{\!_{J}}\tp,\dots,\vect{0}_{\!_{J}}\tp,\vect{\gamma}_{\!_{1}}\tp,\vect{0}_{\!_{J}}\tp,\dots,\vect{0}_{\!_{J}}\tp)\tp, \\
\matr{Q}_g(1,2) {\vect{\theta}} 
& =(\vect{0}_{\!_{J}}\tp,\vect{0}_{\!_{J}}\tp,\vect{\beta}_{\!_{2}}\tp,\vect{0}_{\!_{J}}\tp,\dots,\vect{0}_{\!_{J}}\tp)\tp, &\matr{Q}_g(2,2) {\vect{\theta}}
&=(\vect{0}_{\!_{J}}\tp,\dots,\vect{0}_{\!_{J}}\tp,\vect{0}_{\!_{J}}\tp,\vect{\gamma}_{\!_{2}}\tp,\vect{0}_{\!_{J}}\tp,\dots,\vect{0}_{\!_{J}}\tp)\tp, \\
& \vdots &\vdots&\\
\matr{Q}_g(1,K) {\vect{\theta}} &=(\vect{0}_{\!_{J}}\tp,\dots,\vect{0}_{\!_{J}}\tp,\vect{\beta}_{\!_{K}}\tp,\vect{0}_{\!_{J}}\tp,\dots,\vect{0}_{\!_{J}}\tp)\tp, 
& \matr{Q}_g(2,K) {\vect{\theta}}
& =(\vect{0}_{\!_{J}}\tp,\dots,\vect{0}_{\!_{J}}\tp,\vect{\gamma}_{\!_{K}}\tp)\tp. 
\end{align*}

\section{Proofs}\label{app_proofs}
In this Appendix we prove the results in Lemma~\ref{lem:exp} and Proposition~\ref{prop:discrete} (see Section~\ref{Sec:corr}).  
\subsection{Proof of Lemma~\ref{lem:exp}}
\noindent Notice that the expectation of the periodogram in equation \eqref{periodogram} of the observations $\{\varepsilon_{\!_{i}}, \text{ } i\in\mathcal{I}\}$ is
$$\mean{I_{\,\!{\varepsilon}}(\lambda)} = \sum_{k \in \mathcal{I} } \sum_{j \in \mathcal{I} } \mean{ \varepsilon_{\!_{k}} \varepsilon_{\!_{j}} } \exp(i \lambda [t_k - t_j]).$$
If we replace the expectation $\mean{\varepsilon_{\!_{k}} \varepsilon_{\!_{j}}}$ with the right-hand-side of equation \eqref{autocovar_disc}, we obtain
\begin{align}
\mean{I_{\,\!{\varepsilon}}(\lambda)} &= \frac{2\pi}{N_{\,\!\mathcal{I}}} \sum^{N_{\,\!\mathcal{I}}}_{j=1} P_{\!{\varepsilon}}(\omega_{\!{j}}) \sum_{k \in \mathcal{I}} \sum_{j \in \mathcal{I}}  \exp(i  [\lambda-\omega_{\!{j}}] [t_k-t_j])= \frac{2\pi}{N_{\,\!\mathcal{I}}} \sum^{N_{\,\!\mathcal{I}}}_{j=1} P_{\!{\varepsilon}}(\omega_{\!{j}}) W_{\!{\varepsilon}}(\lambda-\omega_{\!{j}})=\frac{2\pi}{N_{\,\!\mathcal{I}}} P_{\!{\varepsilon}}(\lambda) * W_{\!{\varepsilon}}(\lambda).
\label{exp_perio_disc-app}
\end{align}
\subsection{Proof of Proposition~\ref{prop:discrete}}
Let $\mathcal{F} \{ g_{\!_{j}} \}[k]$ denote the Discrete Fourier Transform (DFT) of the sequence of $m$ numbers $g_{\!_{1}},\dots,g_{\!_{m}}$ into another sequence $h_{\!_{1}},\dots,h_{\!_{m}}$, that is,
\vspace{-.3cm}
\begin{equation*}\label{CFT}
h_{\!_{k}}=\mathcal{F} \{ g_{\!_{j}} \} [k]=\sum^m_{j=1} g_{\!_{j}} \exp(-i k j 2\pi / m), \quad k=1,\dots,m,
\end{equation*}
and $\mathcal{F}^{-1} \{ h_{\!_{k}} \}[j]$ denote the Inverse DFT of the sequence $h_{\!_{1}},\dots,h_{\!_{m}}$ into another sequence $g_{\!_{1}},\dots,g_{\!_{m}}$, that is,
\begin{equation*}\label{inv_CFT}
g_{\!_{j}}=\mathcal{F}^{-1} \{ h_{\!_{k}} \} [j]=\frac{1}{m} \sum^m_{k=1} h_{\!_{k}} \exp(i k j 2\pi / m), \quad j=1,\dots,m.
\end{equation*}
Let $\mathcal{F} \{ g_{\!_{j}} \} [k]$ and $\mathcal{F} \{ \ell_{\!_{j}} \} [k]$ be, respectively, the DFTs of the sequences $\{ g_{\!_{j}} \}$ and $\{ \ell_{\!_{j}} \}$ into the sequences $\{ h_{\!_{k}} \}$ and $\{ m_{\!_{k}} \}$.  Then, the Convolution Theorem states that 
\vspace{-.3cm}
\begin{equation}\label{CONV_theo}
\mathcal{F} \{ g_{\!_{j}} * \ell_{\!_{j}} \} [k] =\mathcal{F} \{ g_{\!_{j}} \} [k] \mathcal{F} \{ \ell_{\!_{j}} \} [k].
\end{equation}

Applying the Convolution Theorem in equation \eqref{CONV_theo} to equation \eqref{exp_perio_disc-app}, we obtain
$$\mathcal{F} \{ P_{\!{\varepsilon}}(\lambda_{j}) \}[k] = \frac{N_{\,\!\mathcal{I}}}{2\pi} \frac{ \mathcal{F} \{ \mean{I_{\,\!{\varepsilon}}(\lambda_{j})} \} [k] } { \mathcal{F} \{W_{\!{\varepsilon}}(\lambda_{j})\}[k] }.$$
The Inverse DFT of the last equation gives the result in equation \eqref{spectral_density_hat-text}.

\bigskip

\newpage

\bibliography{modulated-stars_references}

\begin{thebibliography}{}
\expandafter\ifx\csname natexlab\endcsname\relax\def\natexlab#1{#1}\fi
\providecommand{\url}[1]{\href{#1}{#1}}
\providecommand{\dodoi}[1]{doi:~\href{http://doi.org/#1}{\nolinkurl{#1}}}
\providecommand{\doeprint}[1]{\href{http://ascl.net/#1}{\nolinkurl{http://ascl.net/#1}}}
\providecommand{\doarXiv}[1]{\href{https://arxiv.org/abs/#1}{\nolinkurl{https://arxiv.org/abs/#1}}}

\bibitem[{{Benk{\H{o}}}(2018)}]{Benko2018}
{Benk{\H{o}}}, J.~M. 2018, \mnras, 473, 412, \dodoi{10.1093/mnras/stx2338}

\bibitem[{{Benk{\H{o}}} {et~al.}(2014){Benk{\H{o}}}, {Plachy}, {Szab{\'o}},
  {Moln{\'a}r}, \& {Koll{\'a}th}}]{Benko2014}
{Benk{\H{o}}}, J.~M., {Plachy}, E., {Szab{\'o}}, R., {Moln{\'a}r}, L., \&
  {Koll{\'a}th}, Z. 2014, \apjs, 213, 31, \dodoi{10.1088/0067-0049/213/2/31}

\bibitem[{{Benk{\H{o}}} {et~al.}(2011){Benk{\H{o}}}, {Szab{\'o}}, \&
  {Papar{\'o}}}]{Benko2011}
{Benk{\H{o}}}, J.~M., {Szab{\'o}}, R., \& {Papar{\'o}}, M. 2011, \mnras, 417,
  974, \dodoi{10.1111/j.1365-2966.2011.19313.x}

\bibitem[{Benk{\H{o}} {et~al.}(2010)Benk{\H{o}}, Kolenberg, Szab\'o, Kurtz,
  Bryson, Bregman, Still, Smolec, Nuspl, Nemec, Moskalik, Kopacki, Koll\'ath,
  Guggenberger, Di~Criscienzo, Christensen-Dalsgaard, Kjeldsen, Borucki, Koch,
  Jenkins, \& Van~Cleve}]{Benko2010}
Benk{\H{o}}, J.~M., Kolenberg, K., Szab\'o, R., {et~al.} 2010, \mnras, 409,
  1585

\bibitem[{{Bla{\v{z}}ko}(1907)}]{sb07}
{Bla{\v{z}}ko}, S. 1907, Astronomische Nachrichten, 175, 325,
  \dodoi{10.1002/asna.19071752002}

\bibitem[{Brockwell \& Davis(2016)}]{brockwell2016}
Brockwell, P., \& Davis, R. 2016, Introduction to Time Series and Forecasting,
  2nd edn., Springer Texts in Statistics (Springer International Publishing)

\bibitem[{{Buchler} \& {Koll{\'a}th}(2011)}]{bk11}
{Buchler}, J.~R., \& {Koll{\'a}th}, Z. 2011, \apj, 731, 24,
  \dodoi{10.1088/0004-637X/731/1/24}

\bibitem[{{Catelan} \& {Smith}(2015)}]{cs15}
{Catelan}, M., \& {Smith}, H.~A. 2015, {Pulsating Stars} (Wiley)

\bibitem[{Chadid {et~al.}(2010)Chadid, {{Benk{\H{o}}}, J.M.}, {Szab\'o, R.},
  {Papar\'o, M.}, {Chapellier, E.}, {Kolenberg, K.}, {Poretti, E.}, {Bono, G.},
  {Le Borgne, J.-F.}, {Trinquet, H.}, {Artemenko, S.}, {Auvergne, M.}, {Baglin,
  A.}, {Debosscher, J.}, {Grankin, K. N.}, {Guggenberger, E.}, \& {Weiss, W.
  W.}}]{Chadid2010}
Chadid, M., {{Benk{\H{o}}}, J.M.}, {Szab\'o, R.}, {et~al.} 2010, A\&A, 510, A39

\bibitem[{Chattopadhyay(2017)}]{Chattopadhyay2017}
Chattopadhyay, A.~K. 2017, Incomplete Data in Astrostatistics (American Cancer
  Society), 1--12, \dodoi{https://doi.org/10.1002/9781118445112.stat07942}

\bibitem[{Dahlhaus(1996)}]{D96}
Dahlhaus, R. 1996, Stochastic Processes and their Applications, 62, 139

\bibitem[{Dahlhaus(1997)}]{D97}
---. 1997, The Annals of Statistics, 25, 1, \dodoi{10.1214/aos/1034276620}

\bibitem[{{de Boor}(1978)}]{boor1978}
{de Boor}, C. 1978, {A practical guide to splines} (Springer)

\bibitem[{{Deeming}(1975)}]{Deeming1975}
{Deeming}, T.~J. 1975, \apss, 36, 137, \dodoi{10.1007/BF00681947}

\bibitem[{Eilers {et~al.}(2008)Eilers, Gampe, Marx, \& Rau}]{Eilers2008}
Eilers, P. H.~C., Gampe, J., Marx, B.~D., \& Rau, R. 2008, Statistics in
  Medicine, 27, 3430, \dodoi{https://doi.org/10.1002/sim.3188}

\bibitem[{Eilers \& Marx(1996)}]{Eilers1996}
Eilers, P. H.~C., \& Marx, B.~D. 1996, Statistical Science, 11, 89,
  \dodoi{10.1214/ss/1038425655}

\bibitem[{Elzhov {et~al.}(2016)Elzhov, Mullen, Spiess, \& Bolker}]{minpack2016}
Elzhov, T.~V., Mullen, K.~M., Spiess, A.-N., \& Bolker, B. 2016, minpack.lm: R
  Interface to the Levenberg-Marquardt Nonlinear Least-Squares Algorithm Found
  in MINPACK, Plus Support for Bounds.
\newblock \url{https://CRAN.R-project.org/package=minpack.lm}

\bibitem[{{Feigelson} {et~al.}(2018){Feigelson}, {Babu}, \&
  {Caceres}}]{FBC2018}
{Feigelson}, E.~D., {Babu}, G.~J., \& {Caceres}, G.~A. 2018, Frontiers in
  Physics, 6, 80, \dodoi{10.3389/fphy.2018.00080}

\bibitem[{Gama(2016)}]{NISTunits2016}
Gama, J. 2016, NISTunits: Fundamental Physical Constants and Unit Conversions
  from NIST.
\newblock \url{https://CRAN.R-project.org/package=NISTunits}

\bibitem[{{Gillet}(2013)}]{dg13}
{Gillet}, D. 2013, \aap, 554, A46, \dodoi{10.1051/0004-6361/201220840}

\bibitem[{{Gillet} {et~al.}(2019){Gillet}, {Mauclaire}, {Lemoult}, {Mathias},
  {Devaux}, {de France}, \& {Garrel}}]{gk16}
{Gillet}, D., {Mauclaire}, B., {Lemoult}, T., {et~al.} 2019, \aap, 623, A109,
  \dodoi{10.1051/0004-6361/201833869}

\bibitem[{Hastie {et~al.}(2004)Hastie, Tibshirani, Friedman, \&
  Franklin}]{Hastie2004}
Hastie, T., Tibshirani, R., Friedman, J., \& Franklin, J. 2004, Math. Intell.,
  27, 83, \dodoi{10.1007/BF02985802}

\bibitem[{{Kelly} {et~al.}(2014){Kelly}, {Becker}, {Sobolewska},
  {Siemiginowska}, \& {Uttley}}]{Kelly2014}
{Kelly}, B.~C., {Becker}, A.~C., {Sobolewska}, M., {Siemiginowska}, A., \&
  {Uttley}, P. 2014, \apj, 788, 33, \dodoi{10.1088/0004-637X/788/1/33}

\bibitem[{{Koch} {et~al.}(2010){Koch}, {Borucki}, {Basri}, {Batalha}, {Brown},
  {Caldwell}, {Christensen-Dalsgaard}, {Cochran}, {DeVore}, {Dunham},
  {Gautier}, {Geary}, {Gilliland}, {Gould}, {Jenkins}, {Kondo}, {Latham},
  {Lissauer}, {Marcy}, {Monet}, {Sasselov}, {Boss}, {Brownlee}, {Caldwell},
  {Dupree}, {Howell}, {Kjeldsen}, {Meibom}, {Morrison}, {Owen}, {Reitsema},
  {Tarter}, {Bryson}, {Dotson}, {Gazis}, {Haas}, {Kolodziejczak}, {Rowe}, {Van
  Cleve}, {Allen}, {Chandrasekaran}, {Clarke}, {Li}, {Quintana}, {Tenenbaum},
  {Twicken}, \& {Wu}}]{Koch2010}
{Koch}, D.~G., {Borucki}, W.~J., {Basri}, G., {et~al.} 2010, \apjl, 713, L79,
  \dodoi{10.1088/2041-8205/713/2/L79}

\bibitem[{Lomb(1976)}]{Lomb1976}
Lomb, N.~R. 1976, Astrophysics and Space Science, 39, 447

\bibitem[{{Netzel} {et~al.}(2018){Netzel}, {Smolec}, {Soszy{\'n}ski}, \&
  {Udalski}}]{hnea18}
{Netzel}, H., {Smolec}, R., {Soszy{\'n}ski}, I., \& {Udalski}, A. 2018, \mnras,
  480, 1229, \dodoi{10.1093/mnras/sty1883}

\bibitem[{{Plachy} {et~al.}(2019){Plachy}, {Moln{\'a}r}, {B{\'o}di}, {Skarka},
  {Szab{\'o}}, {Szab{\'o}}, {Klagyivik}, {S{\'o}dor}, \& {Pope}}]{epea19}
{Plachy}, E., {Moln{\'a}r}, L., {B{\'o}di}, A., {et~al.} 2019, \apjs, 244, 32,
  \dodoi{10.3847/1538-4365/ab4132}

\bibitem[{Poretti {et~al.}(2010)Poretti, {Papar\'o, M.}, {Deleuil, M.},
  {Chadid, M.}, {Kolenberg, K.}, {Szab\'o, R.}, {Benk"o, J. M.}, {Chapellier,
  E.}, {Guggenberger, E.}, {Le Borgne, J. F.}, {Rostagni, F.}, {Trinquet, H.},
  {Auvergne, M.}, {Baglin, A.}, {Sarro, L. M.}, \& {Weiss, W.
  W.}}]{Poretti2010}
Poretti, E., {Papar\'o, M.}, {Deleuil, M.}, {et~al.} 2010, A\&A, 520, A108

\bibitem[{Priestley(1981)}]{Priestley1981}
Priestley, M.~B. 1981, Spectral analysis and time series (Academic Press)

\bibitem[{{R Core Team}(2021)}]{R2021}
{R Core Team}. 2021, R: A Language and Environment for Statistical Computing, R
  Foundation for Statistical Computing, Vienna, Austria.
\newblock \url{https://www.R-project.org}

\bibitem[{{Richards} {et~al.}(2011){Richards}, {Starr}, {Butler}, {Bloom},
  {Brewer}, {Crellin-Quick}, {Higgins}, {Kennedy}, \&
  {Rischard}}]{Richards_2011}
{Richards}, J.~W., {Starr}, D.~L., {Butler}, N.~R., {et~al.} 2011, \apj, 733,
  10, \dodoi{10.1088/0004-637X/733/1/10}

\bibitem[{{Smith}(1995)}]{hs95}
{Smith}, H.~A. 1995, Cambridge Astrophysics Series, 27

\bibitem[{{S{\'o}dor} {et~al.}(2012){S{\'o}dor}, {Hajdu}, {Jurcsik}, {Szeidl},
  {Posztob{\'a}nyi}, {Hurta}, {Belucz}, \& {Kun}}]{Sodor2012}
{S{\'o}dor}, {\'A}., {Hajdu}, G., {Jurcsik}, J., {et~al.} 2012, \mnras, 427,
  1517, \dodoi{10.1111/j.1365-2966.2012.21837.x}

\bibitem[{{Stothers}(2006)}]{rs06}
{Stothers}, R.~B. 2006, \apj, 652, 643, \dodoi{10.1086/508135}

\bibitem[{{Wong} {et~al.}(2015){Wong}, {Kashyap}, {Lee}, \& {van
  Dyk}}]{Wong2016}
{Wong}, R. K.~W., {Kashyap}, V.~L., {Lee}, T. C.~M., \& {van Dyk}, D.~A. 2015,
  arXiv e-prints, arXiv:1508.07083.
\newblock \doarXiv{1508.07083}

\bibitem[{{Xu} {et~al.}(2021){Xu}, {G{\"u}nther}, {Kashyap}, {Lee}, \&
  {Zezas}}]{Xu2021}
{Xu}, C., {G{\"u}nther}, H.~M., {Kashyap}, V.~L., {Lee}, T. C.~M., \& {Zezas},
  A. 2021, \aj, 161, 184, \dodoi{10.3847/1538-3881/abe0b6}

\bibitem[{Zhou {et~al.}(1998)Zhou, Shen, \& Wolfe}]{Zhou1998}
Zhou, S., Shen, X., \& Wolfe, D.~A. 1998, The Annals of Statistics, 26, 1760

\end{thebibliography}
\bibliographystyle{aasjournal}

\end{document}